\documentclass[a4paper, amsfonts, amssymb, amsmath, reprint, showkeys, nofootinbib, oneside, superscriptaddress, preprintnumbers]{revtex4-1}
\usepackage[english]{babel}
\usepackage[utf8]{inputenc}
\usepackage{slashed}
\usepackage{lipsum} 
\usepackage[force]{feynmp-auto}
\usepackage{graphicx}
\usepackage{dcolumn}
\usepackage{bm}
\usepackage{color}
\usepackage{multirow}
\usepackage{textgreek}
\usepackage{tabularx}
\newcommand{\pq}{\bar{P}\cdot\bar{q}}
\usepackage{multirow, makecell}
\newcolumntype{C}{>{\centering\arraybackslash}X}
\usepackage[colorinlistoftodos, color=green!40, prependcaption]{todonotes}
\usepackage{amsthm}
\usepackage{mathtools}
\usepackage{physics}
\usepackage{xcolor}
\usepackage{graphicx}
\usepackage[left=23mm,right=13mm,top=35mm,columnsep=15pt]{geometry} 
\usepackage{adjustbox}
\usepackage{placeins}
\usepackage[T1]{fontenc}
\usepackage{lipsum}
\usepackage{csquotes}

\usepackage[pdftex, pdftitle={Article}, pdfauthor={Author}]{hyperref} % For hyperlinks in the PDF
\DeclareUnicodeCharacter{2009}{\,} 

\renewcommand{\>}{\rangle}
\newcommand{\<}{\langle}
\newcommand{\bomega}{\bar{\omega}}
\newcommand{\cev}[1]{\reflectbox{\ensuremath{\vec{\reflectbox{\ensuremath{#1}}}}}}

\begin{document}
\title{Generalised parton distributions from the off-forward Compton amplitude in lattice QCD}

\author{A.~Hannaford-Gunn}
    \affiliation{CSSM, Department of Physics, The University of Adelaide, Adelaide SA 5005, Australia}
\author{K.~U.~Can}
    \affiliation{CSSM, Department of Physics, The University of Adelaide, Adelaide SA 5005, Australia}
    \author{R.~Horsley}
    \affiliation{School of Physics, University of Edinburgh, Edinburgh EH9 3JZ, UK}
    \author{Y.~Nakamura}
    \affiliation{RIKEN Center for Computational Science, Kobe, Hyogo 650-0047, Japan}
    \author{H.~Perlt}
    \affiliation{Insitut f\"ur Theoretische Physik, Universist\"at Leipzig, 04103 Leipzig, Germany}
\author{P.~E.~L.~Rakow}
    \affiliation{Theoretical Physics Division, Department of Mathematical Sciences, University of Liverpool, Liverpool L69 3BX, UK}
    \author{G.~Schierholz}
    \affiliation{Deutsches Elektronen-Synchrotron DESY, Notkestr.~85, 22607 Hamburg, Germany}
      \author{H.~St\"uben}
    \affiliation{Regionales Rechenzentrum, Universit\"at Hamburg, 20146 Hamburg, Germany}
    \author{R.~D.~Young}
    \affiliation{CSSM, Department of Physics, The University of Adelaide, Adelaide SA 5005, Australia}
    \author{J.~M.~Zanotti}
    \affiliation{CSSM, Department of Physics, The University of Adelaide, Adelaide SA 5005, Australia}

    \collaboration{CSSM/QCDSF/UKQCD Collaborations}
	\noaffiliation

\date{\today} % Leave empty to omit a date

\begin{abstract}
We determine the properties of generalised parton distributions (GPDs) from a lattice QCD calculation of the off-forward Compton amplitude (OFCA). By extending the Feynman-Hellmann relation to second-order matrix elements at off-forward kinematics, this amplitude can be calculated from lattice propagators computed in the presence of a background field. Using an operator product expansion, we show that the deeply-virtual part of the OFCA can be parameterised in terms of the low-order Mellin moments of the GPDs. We apply this formalism to a numerical investigation for zero-skewness kinematics at two values of the soft momentum transfer, $t = -1.1, -2.2 \;\text{GeV}^2$, and a pion mass of $m_{\pi}\approx 470\;\text{MeV}$. The form factors of the lowest two moments of the nucleon GPDs are determined, including the first lattice QCD determination of the $n=4$ moments. Hence we demonstrate the viability of this method to calculate the OFCA from first principles, and thereby provide novel constraint on the $x$- and $t$-dependence of GPDs.
\end{abstract}

\keywords{Compton scattering, generalised parton distributions, lattice QCD, operator product expansion, Feynman-Hellmann}
\preprint{ADP-21-15/T1162}
\preprint{DESY-21-167}
\preprint{Liverpool LTH 1271}
\maketitle

\section{Introduction}
\label{sec:intro}

Since the 1990s, generalised parton distributions (GPDs) have been recognised as crucial observables in understanding hadron structure \cite{mullerscaling, jiog, radyushkinscaling}. They encode the spatial distribution of quarks and gluons in a fast-moving hadron \cite{burkardt}. Moreover, their Mellin moments contain information about the spin and orbital angular momentum of hadron constituents \cite{jiog}, which would resolve the decades-old `proton spin puzzle' \cite{thomasprotonspin, bassprotonspin}. Finally, more recent research has explored the relationship between GPDs and `mechanical' properties: pressure, energy and force distributions within hadrons \cite{mechanprops2, dtermexp}. 

%A recent Nature letter reported on the experimental measurement of GPDs to determine the pressure distribution within a proton \cite{dtermexp}.

GPDs can be measured from off-forward Compton scattering processes, such as deeply virtual Compton scattering (DVCS), which have been carried out at HERA \cite{heradvcs1, heradvcs2, heradvcs3, heradvcs4, heradvcs5}, COMPASS \cite{compassdvcs}, JLab \cite{jlabdvcs1, jlabdvcs2, jlabdvcs3, jlabdvcs4}, and are planned to be carried out in the future at the electron-ion collider \cite{eic}. However, due to the high dimensionality of GPDs, they are difficult to extract directly from experiment, and global fits require assumptions about their functional form \cite{gpdphenomreview, gpdphenomreview2}. Therefore, a stronger theoretical understanding of GPD behaviour would allow for more precise experimental determinations.

Historically, lattice QCD calculations have been limited to Mellin moments of GPDs from matrix elements of leading-twist local operators \cite{gpdlatt1, gpdlatt2, gpdlatt3, gpdlatt4, gpdlatt5, gpdlatt6, gpdlatt7, gpdlatt8, gpdlatt9, gpdlatt10, detmold_shanhan_latt1, detmold_shanhan_latt2}. However, it has long been known that matrix elements of leading-twist suffer from power-divergent renormalisation due to the broken Lorentz symmetry on the lattice \cite{martinelli_power_div}. For the lowest moments, this can be controlled \cite{opmixing1}, but it becomes more difficult for higher moments \cite{opmixing2}. As such, the $n=3$ moments are the highest so far computed \cite{gpdlatt6}. Determinations of higher moments would allow for better constraint of GPDs \cite{momentstopds, detmold_mom_reconstruction}.

More recently, there have also been major efforts to reconstruct the full $x$-dependence of parton distributions in lattice QCD, using the pseudo- \cite{radpseudo} and quasi-distribution \cite{jiquasi} methods---see Refs.~\cite{xdependence_review, latticewhitepaper} for reviews. This includes recent calculations of quasi-GPDs \cite{pionquasi, nucleonquasi1, nucleonquasi2}. These methods aim to extract the light-cone distributions directly, whereas in the present work we are interested in the Compton scattering amplitude, from which GPDs may be accessed experimentally.

%However, the matching procedure to recover light-cone distributions from quasi-distributions may introduce significant systematic uncertainties \cite{xdependence_review}. Moreover, the renormalisation of quasi-distributions requires the removal of power-divergent terms, which introduces systematic errors and complicates the continuum extrapolation \cite{ latticewhitepaper}.

 In this paper, we determine properties of GPDs from a calculation of the off-forward Compton amplitude (OFCA) in lattice QCD. The OFCA is defined as
\begin{equation}
       T^{\mu\nu}\equiv i\int d^4ze^{\frac{i}{2}(q+q')\cdot z} \<P'|T\{j^{\mu}(z/2)j^{\nu}(-z/2)\}|P\>,
       \label{vcadef}
   \end{equation}
   and describes the process of $\gamma^{*} (q)N(P)  \to \gamma^{*}(q')N(P') $, with $q_{\mu}\neq 0 \neq q'_{\mu}$. Here, $j^{\mu}$ is the hadronic vector current, and we limit ourselves to the case where the scattered hadron is a nucleon. 
   
   Besides GPDs, this amplitude gives access to a range of interesting physical quantities, including generalised polarisabilities \cite{gpols0, pasquini_DRs1, gpols1, gpols2} and the subtraction function \cite{offfwdsub1, offfwdsub2}. In the high energy region ($|q^2|$ and/or $|q'^2|\gg \Lambda_{\text{QCD}}^2$), it is dominated by contributions from GPDs.
   
   By calculating the Compton amplitude, we overcome the issues of power-divergent renormalisation that the leading-twist matrix elements suffer from \cite{opewope1, opewope2}. Moreover, with the correction for lattice systematics, our calculation contains the same higher-twist contributions as the physical amplitude, which are of interest beyond their connection to leading-twist GPDs \cite{aslantwist3}. Therefore, the present calculation bears many similarities to the hadronic tensor approach, which aims to access the forward structure functions from the direct calculation of four-point functions \cite{latthadronic1, towardslatticehadronic}.

\begin{figure}[t!]
\centering
\includegraphics[width=0.64\linewidth]{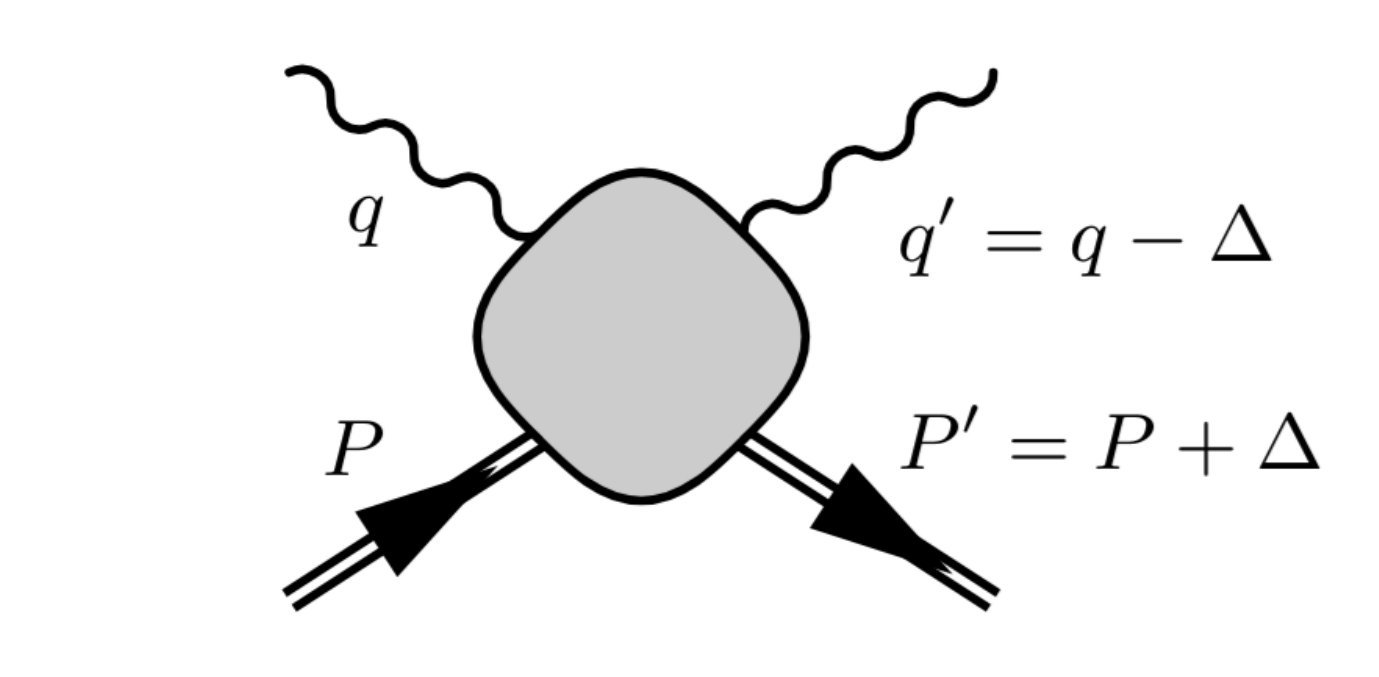}
%\begin{fmffile}{123454}
%\begin{fmfgraph*}(100,60)
%        \fmfkeep{photon}
%        \fmfleft{i1,i2}
%        \fmfright{o1,o2}
%       \fmfpoly{smooth,filled=20, pull=1.6}{v4,v3,v2,v1}
%         \fmf{dbl_plain_arrow,label=$P$, label.side=left, width=1.0}{i1,v1}
%          \fmf{dbl_plain_arrow,label=$P'=P+\Delta$, label.side=left, width=1.0}{v4,o1}
%        \fmf{photon,label=$q$,label.side=right}{i2,v2}
%        \fmf{photon,label=$q'=q-\Delta$,label.side=right}{v3,o2}
%    \end{fmfgraph*}
%\end{fmffile}
    \caption{The Feynman diagram for off-forward $\gamma^{*} (q)N(P)  \to \gamma^{*}(q')N(P') $ scattering.}
    \label{offfwdpic}
    \end{figure}

The method presented here to calculate the OFCA is an extension of Feynman-Hellmann methods used previously to determine the \textit{forward} Compton amplitude \cite{fwdletter, fwdpaper}. Two-point correlators calculated in the presence of a weakly-coupled background field field can be expanded in powers of the coupling, with their second-order contribution in terms of four-point functions. As such, Feynman-Hellmann methods are a feasible alternative to the direct calculation of four-point functions.

The numerical results presented here are the first lattice QCD determination of the off-forward Compton amplitude. This calculation is performed at the SU(3) flavour symmetric point and larger-than-physical pion mass \cite{configs}, for two values of the soft momentum transfer, $t=-1.1, -2.2 \;\text{GeV}^2$, with zero-skewness kinematics. In this preliminary work, we assume leading-twist dominance, since our hard scale is in the perturbative region: $\bar{Q}^2\approx 6-7 \;\text{GeV}^2$. As such, we fit Mellin moments of the OFCA, and interpret these as the moments of GPDs. 

The structure of this paper follows: in section \ref{sec:bckgrnd} we review key properties of the OFCA; in section \ref{sec:FHrel} we derive the Feynman-Hellmann relation that allows us to determine the OFCA; in section \ref{sec:gca} we use an operator product expansion to parameterise the scalar amplitudes of the OFCA in terms of GPD moments; in section \ref{sec:numcalc} we outline the details of our numerical calculation; and finally in section \ref{sec:results} we present our results.

\section{Background} \label{sec:bckgrnd}

We start by considering a general process of off-forward photon-nucleon scattering: $\gamma^{*}(q)N(P)\to \gamma^{*}(q')N(P')$ (see Figure \ref{offfwdpic}). 

We choose the basis of momentum vectors
\begin{equation*}
         \bar{P}= \frac{1}{2}(P+P'), \quad \bar{q} = \frac{1}{2}(q+q'), \quad \Delta = P' - P = q-q'.
\end{equation*}
From these, we can form at most four linearly independent scalar variables: two scaling variables,
 \begin{equation*}
    \bomega=\frac{2\bar{P}\cdot\bar{q}}{\bar{Q}^2}, \quad  \vartheta=-\frac{\Delta\cdot\bar{q}}{\bar{Q}^2},
\end{equation*}
and the soft and hard momentum transfers, respectively,
\begin{equation*}
    t = \Delta^2, \quad \bar{Q}^2 = -\bar{q}^2.
\end{equation*}
In terms of these scalars, the usual skewness variable \cite{jiofpds} is $\xi = \vartheta /\bomega$, and hence $\vartheta=0$, $\bomega\neq0$ implies that $\xi=0$. In terms of the conventional deeply virtual Compton scattering (DVCS) kinematics, where $q'^2=0$, we have that $\vartheta\simeq 1$ and $\bomega\simeq\xi^{-1}$ for large $-q^2$.

\subsection*{Tensor Decomposition}

The amplitude for this process, the off-forward Compton amplitude (OFCA), is defined in Eq.~\eqref{vcadef}. It can be decomposed into 18 linearly independent tensor structures \cite{perrottet, tarrach, gpols0, eichmannfischer, dv2quasireal}:
\begin{equation}
     T^{\mu\nu}(\bomega, \vartheta, t, \bar{Q}^2) = \sum _{i=1}^{18} \mathcal{A}_i(\bomega, \vartheta, t, \bar{Q}^2) L^{\mu\nu}_i,
     \label{startingpt}
\end{equation}
where $\mathcal{A}_i$ are invariant amplitudes and $L^{\mu\nu}_i$ are Lorentz tensors and Dirac bilinears. 

As a consequence of the Ward identities of the OFCA, $q_{\mu}T^{\mu\nu}=0=q'_{\nu}T^{\mu\nu}$, contributions to the Compton amplitude that are proportional to $q_{\nu}$ or $q'_{\mu}$ are not linearly independent. Hence we can write the OFCA as
\begin{equation*}
    T^{\mu\nu}=\bar{T}_{\rho\sigma}\mathcal{P}^{\mu\rho}\mathcal{P}^{\sigma\nu},
\end{equation*}
where the gauge projector is
\begin{equation}
     \mathcal{P}^{\mu\nu} = g^{\mu\nu}-\frac{q'^{\mu}q^{\nu}}{q\cdot q'},
     \label{gaugeproj}
\end{equation}
and $\bar{T}_{\mu\nu}$ is the OFCA with no $q_{\nu}$ or $q'_{\mu}$ terms. 

We will choose a basis for the tensor decomposition of $\bar{T}_{\mu\nu}$, since all other terms are entirely determined by the Ward identities. In our chosen basis, the OFCA (before gauge projection) is
\begin{widetext}
\begin{equation}
    \begin{split}
        \bar{T}_{\mu\nu} & = \frac{1}{2\pq}\bigg[-\Big ( h\cdot \bar{q} \mathcal{H}_1+ e\cdot \bar{q} \mathcal{E}_1  \Big)g_{\mu\nu} + \frac{1}{\pq}\Big ( h\cdot \bar{q} \mathcal{H}_2+ e\cdot \bar{q} \mathcal{E}_2  \Big)\bar{P}_{\mu}\bar{P}_{\nu} + \mathcal{H}_3 h_{\{\mu}\bar{P}_{\nu\}}\bigg ]
        \\ & + \frac{i}{2\pq}\epsilon_{\mu\nu\rho\kappa}\bar{q}^{\rho}\Big ( \tilde{h}^{\kappa}  \tilde{\mathcal{H}}_1+ \tilde{e}^{\kappa} \tilde{\mathcal{E}}_1  \Big)+ \frac{i}{2(\pq)^2}\epsilon_{\mu\nu\rho\kappa}\bar{q}^{\rho}\Big [ \big (\pq \tilde{h}^{\kappa}-\tilde{h}\cdot \bar{q}\bar{P}^{\kappa}\big) \tilde{\mathcal{H}}_2+ \big (\pq \tilde{e}^{\kappa}-\tilde{e}\cdot \bar{q}\bar{P}^{\kappa}\big) \tilde{\mathcal{E}}_2  \Big]
        \\ & + \Big(\bar{P}_{\mu}q'_{\nu}+\bar{P}_{\nu}q_{\mu}\Big) \Big( h\cdot \bar{q} \mathcal{K}_1+ e\cdot \bar{q} \mathcal{K}_2 \Big)   + \Big(\bar{P}_{\mu}q'_{\nu}-\bar{P}_{\nu}q_{\mu}\Big) \Big( h\cdot \bar{q} \mathcal{K}_3+ e\cdot \bar{q} \mathcal{K}_4 \Big) + q_{\mu}q'_{\nu}\big(h\cdot\bar{q}-e\cdot\bar{q}\big)\mathcal{K}_5 \\ & + h_{[\mu}\bar{P}_{\nu]}\mathcal{K}_6 + \Big(h_{\mu}q'_{\nu}+h_{\nu}q_{\mu}\Big)\mathcal{K}_7+ \Big(h_{\mu}q'_{\nu}-h_{\nu}q_{\mu}\Big)\mathcal{K}_8 + \bar{P}_{\{\mu}\bar{u}(P')i\sigma_{\nu\}\alpha}u(P)\bar{q}^{\alpha} \mathcal{K}_9,
        \label{expltensordecomp}
    \end{split}
\end{equation}
 \end{widetext}
 where we have introduced the Dirac bilinears
   \begin{equation}
   \begin{split}
       h^{\mu} & = \bar{u}'\gamma ^{\mu}u, \quad e^{\mu}=\bar{u}'\frac{i\sigma ^{\mu\alpha}\Delta_{\alpha}}{2m_N}u,
  \\   \tilde{h}^{\mu} & = \bar{u}'\gamma ^{\mu}\gamma_5u, \quad \tilde{e}^{\mu}= \frac{\Delta^{\mu}}{2m_N}\bar{u}'\gamma_5u.
  \label{bilineardef}
   \end{split}
\end{equation}
In Eq.~\eqref{expltensordecomp}, there are nine $\mathcal{K}$, five unpolarised ($\mathcal{H}$ and $\mathcal{E}$) and four polarised ($\tilde{\mathcal{H}}$ and $\tilde{\mathcal{E}}$) amplitudes, which gives 18 in total.

The basis in Eq.~\eqref{expltensordecomp} is chosen to match onto the high-energy limit, which we will derive in section \ref{sec:gca}.
While this does introduce kinematic singularities into our basis, these are not relevant to the leading-twist contribution or our numerical calculation.

The amplitudes of Eq.~\eqref{expltensordecomp} also reduce in the forward ($t\to0$) limit to the more well-known functions of the forward Compton amplitude:
    \begin{equation*}
        \begin{split}
           & \mathcal{H}_1 \overset{t\to0}{\longrightarrow} \mathcal{F}_1, \quad \mathcal{H}_2 +  \mathcal{H}_3 \overset{t\to0}{\longrightarrow} \mathcal{F}_2, 
           \\ & \quad 
            \tilde{\mathcal{H}}_1 \overset{t\to0}{\longrightarrow} \tilde{g}_1, \quad \tilde{\mathcal{H}}_2 \overset{t\to0}{\longrightarrow} \tilde{g}_2,
        \end{split}
    \end{equation*}
    where $\mathcal{F}_{1,2}$ are the Compton structure functions \cite{fwdpaper} and $\text{Im}\tilde{g}_{1,2} = 2\pi g_{1,2}$, for $g_{1,2}$ the spin-dependent, deep inelastic structure functions \cite{manohar}. On the other hand, the $\mathcal{K}$ amplitudes vanish in the forward limit.

\subsection*{Dispersion Relation}

As in the forward case, we can use the analytic features of the amplitudes in Eq.~\eqref{expltensordecomp} to write out a dispersion relation. For instance, following Refs.~\cite{pasquini_DRs1, pasquini_DRs2}, $\mathcal{H}_1$ and $\mathcal{E}_1$ satisfy subtracted dispersion relations:
\begin{equation}
  \begin{split}
        \mathcal{H}_1(\bomega, \vartheta, t, \bar{Q}^2)& = S_1(\vartheta, t, \bar{Q}^2) +  \overline{\mathcal{H}}_1(\bomega, \vartheta, t, \bar{Q}^2),
      \\    \mathcal{E}_1(\bomega, \vartheta, t, \bar{Q}^2) & = -S_1(\vartheta, t, \bar{Q}^2) +  \overline{\mathcal{E}}_1(\bomega, \vartheta, t, \bar{Q}^2),
      \label{subDR}
  \end{split}
\end{equation}
where we have introduced
\begin{equation*}
    \overline{\mathcal{H}}_1(\bomega, \vartheta, t, \bar{Q}^2) = \frac{2\bomega^2}{\pi}\int_0 ^1 dx \frac{x \text{Im}\mathcal{H}_1(\bomega, \vartheta, t, \bar{Q}^2)}{1-x^2\bomega^2 -i\epsilon},
\end{equation*}
and similarly for $\mathcal{H}_1\to\mathcal{E}_1$. 

The subtraction function in Eq.~\eqref{subDR} is a generalisation of the forward Compton amplitude subtraction function \cite{gasserleutwyler}: $S_1(\vartheta, t, \bar{Q}^2)\overset{t\to0}{\longrightarrow} S_1(Q^2)$, which has been studied elsewhere \cite{offfwdsub1, offfwdsub2}. The amplitudes $\mathcal{H}_{2,3}$ and $\mathcal{E}_2$ require no subtraction in their dispersion relations \cite{pasquini_DRs1, pasquini_DRs2}.

The forward limit of $\overline{\mathcal{H}}_1$ is $$ \overline{\mathcal{H}}_1(\bomega, \vartheta, t, \bar{Q}^2) \overset{t\to0}{\longrightarrow} {4\omega^2}\int_0^{1}dx\frac{xF_1(x, Q^2)}{1-x^2\omega^2-i\epsilon},
    $$
    where $F_1$ is the deep inelastic scattering structure function \cite{pasquini_DRs1}. However, unlike the forward case, there is no optical theorem to relate $\text{Im}\mathcal{H}_{1,2}$ to an inclusive cross section. Instead, these amplitudes can be measured directly by exclusive processes such as DVCS.

\subsection*{Generalised Parton Distributions}

At high energies ($\bar{Q}^2\gg \Lambda_{\text{QCD}}^2$), the amplitudes of Eq.~\eqref{expltensordecomp} are dominated by convolutions of GPDs \cite{jiog, gpdfactorization}:
\begin{equation*}
    \begin{split}
        \mathcal{A} & \simeq \int dx G(x,\vartheta/\bomega,t)\bigg[ \frac{ \bomega}{1 +x\bomega -i\epsilon} \pm  \frac{ \bomega}{1-x\bomega  -i\epsilon} \bigg],
    \end{split}
\end{equation*}
where $G$ is a GPD. Or, in the Euclidean region, $|\bomega|<1$, 
\begin{equation*}
     \mathcal{A} \simeq \sum_n \bomega ^n\int dx x^{n-1} G(x,\vartheta/\bomega,t).
\end{equation*}

Formally, GPDs are defined by the off-forward matrix element of a light-cone operator. For a light-like vector $n^{\mu}$ such that $n\cdot \bar{P}=1$ (and hence $\xi=-n\cdot\Delta/2$) and taking light-cone gauge $n\cdot U=0$, we have \cite{jiog,jidvcs}
\begin{equation}
\begin{split}
       \int \frac{d\lambda}{2\pi}e^{i\lambda x}&\<P' |\bar{\psi}_q(-\lambda n/2)  {\slashed{n}}\psi_q(\lambda n/2)|P\>
      \\ &  =H^q(x,\vartheta/\bomega,t)\bar{u}(P')\gamma^{\mu}n_{\mu}u(P)
      \\ &+E^q(x, \vartheta/\bomega,t)\bar{u}(P')\frac{i\sigma^{\mu\nu}n_{\mu}\Delta_{\nu}}{2m_N}u(P),
        \label{GPDdef}
        \end{split}
\end{equation}
where $H^q$ and $E^q$ are the unpolarised twist-two GPDs for a quark of flavour $q$. 
It is not possible to directly calculate the quantity in Eq.~\eqref{GPDdef} on the lattice, due to the Euclidean signature of spacetime. 

Instead, we can relate GPDs to a basis of leading-twist local operators. These local operators are
\begin{align}
   \begin{split}
    & \mathcal{O}^{(n)\mu_1...\mu_{n}}_q
    =\bar{\psi}_q\gamma^{\{\mu_1}i{\overset{\leftrightarrow} D}^{\mu_2}... i {\overset{\leftrightarrow} D}^{\mu_n\}} \psi_q- \textnormal{traces},
    \label{localtwisttwo}
   \end{split}
\end{align}
where ${\overset{\leftrightarrow} D}= \frac{1}{2}(\vec{D}- \cev{D})$. See appendix \ref{sec:appendixBG} for the symmetrisation convention of the Lorentz indices.

The off-forward nucleon matrix elements of the operators in Eq.~\eqref{localtwisttwo} are \cite{jiofpds}
\begin{widetext}
 \begin{equation}
   \begin{split}
   & \< P' |\mathcal{O}^{(n+1) \kappa\mu_1...\mu_{n}}_q  (0)| P\>  
   = \bar{u}(P')\gamma ^{\{\kappa}u(P)\sum ^{n}_{i=0}A^q_{n+1, i}(t) \Delta ^{\mu_1}...\Delta ^{\mu_{i}}\bar{P}^{\mu_{i+1}}...\bar{P}^{\mu_{n}\}}\\ &+ \bar{u}(P')\frac{i\sigma ^{\{\kappa\alpha}\Delta_{\alpha}}{2m_N}u(P)\sum ^{n}_{i=0}B^q_{n+1, i}(t) \Delta ^{\mu_1}...\Delta ^{\mu_{i}}\bar{P}^{\mu_{i+1}}...\bar{P}^{\mu_{n}\}}+ C^q_{n+1}(t)\text{mod}(n,2)\frac{\bar{u}(P')u(P)}{m_N} \Delta^{\{\kappa}\Delta ^{\mu_1}...\Delta^{\mu_n \}},
    \label{offfwdlocal2}
    \end{split}
\end{equation}
\end{widetext}
where the Lorentz scalars $A_{n,i}^q$, $B_{n,i}^q$ and $C_{n}^q$ are generalised form factors (GFFs).

By Taylor expanding Eq.~\eqref{GPDdef}, one can relate the GFFs from Eq.~\eqref{offfwdlocal2} to the GPDs $H$ and $E$:
\begin{equation}
   \begin{split}
       \int _{-1}^1 dx x^{n} & H^q(x,\vartheta/\bomega,t)
      =\sum^n_{i=0,2,4}(-2\vartheta/\bomega)^iA^q_{n+1,i}(t)\\&+\text{mod}(n,2)(-2\vartheta/\bomega)^{n+1}C^q_{n+1}(t),
  \\ \int _{-1}^1 dx x^{n} & E^q(x,\vartheta/\bomega,t)
  =\sum^n_{i=0,2,4}(-2\vartheta/\bomega)^iB^q_{n+1,i}(t) \\ &-\text{mod}(n,2)(-2\vartheta/\bomega)^{n+1}C^q_{n+1}(t),
   \end{split}
   \label{polynomiality}
\end{equation}
recalling that $\xi = \vartheta/\bomega$ in the scalars defined at the start of this section. These equations are the famous `polynomiality' of GPDs \cite{jiofpds}.

A proof-of-principle determination of GPD moments is the ultimate aim of this paper. Specifically, we will calculate the linear combination of zero-skewness moments,
$$ A^q_{n,0}(t) + \frac{t}{8m_N^2}B^q_{n,0}(t), \quad n=2,4.
$$
Equivalent expressions for polarised GPDs, $\tilde{H}$ and $\tilde{E}$, are given in appendix \ref{sec:appendixBG}.

\section{Feynman-Hellmann Relation} \label{sec:FHrel}

In this section, we will show how to calculate the off-forward Compton amplitude from Feynman-Hellmann methods in lattice QCD. Feynman-Hellmann methods are a subset of background field methods, in which a two-point function is calculated in the presence of a weakly-coupled field or current. This induces perturbations to the two-point function, thereby giving access to observables that may be difficult to calculate with a direct $n$-point function.

For the Feynman-Hellmann derivation presented here, we expand the perturbed propagator by means of a Dyson expansion \cite{mischarogerdyson}. This is used to approximate a derivative of the propagator, similar to Refs.~\cite{detmoldfh, touissaintfh, bourchardfh, changaxialFH}, and hence extract the OFCA for off-forward kinematics. 

This differs from our previous proof of the forward Feynman-Hellmann relation \cite{fwdpaper}, where we expressed the perturbed correlators as $\mathcal{G}_{\lambda}(\tau)\simeq A_{\lambda}e^{-E_{\lambda}\tau}$, and related the derivatives of the perturbed energy, $E_{\lambda}$, to the Compton amplitude. While it is still possible to derive a Feynman-Hellmann relation for the OFCA in terms of derivatives of perturbed energies \cite{alecthesis}, such a proof is made difficult by the fact that degeneracies in the unperturbed spectrum cause there to be two low-lying perturbed energies. Similar considerations are needed for nucleon electromagnetic form factors from Feynman-Hellmann \cite{collabffs}. By contrast, the Dyson expansion and correlator derivative formalism presented below circumvents this difficulty.

We introduce two spatially-oscillating background fields to the QCD lagrangian density:
\begin{equation}
    \begin{split}
         \mathcal{L}_{\text{FH}}(x) &  = \mathcal{L}_{\text{QCD}}(x)  +\lambda_1(e^{i\mathbf{q}_1\cdot\mathbf{x}}+e^{-i\mathbf{q}_1\cdot\mathbf{x}})j_3(x)
        \\ & + \lambda_2(e^{i\mathbf{q}_2\cdot\mathbf{x}}+e^{-i\mathbf{q}_2\cdot\mathbf{x}}) j_3(x),
    \label{pertlang}
    \end{split}
\end{equation}
where $j_3(x)=Z_V \bar{\psi}_q(x)i\gamma_3\psi_q(x)$, and $Z_V$ is the lattice renormalisation constant for a local vector current.

Therefore, the perturbed Hamiltonian is
\begin{equation}
    {H}_{\text{FH}}= {H}_{\text{QCD}} -\sum_k\lambda_kV_k(\tau),
    \label{pertham}
\end{equation} 
where 
\begin{equation*}
     V_k(\tau) = \int d^3 x (e^{i\mathbf{q}_k\cdot\mathbf{x}}+e^{-i\mathbf{q}_k\cdot\mathbf{x}}) j_3(x).
\end{equation*}
Simulating with the perturbed Lagrangian in Eq.~\eqref{pertlang} leads to a modified lattice two-point propagator:
\begin{equation}
     \begin{split}
        &  \mathcal{G}_{\boldsymbol{\lambda}} (\tau, \mathbf{p}')= \Gamma_{\beta\alpha}\int d^3 x e^{-i\mathbf{p}'\cdot\mathbf{x}}\prescript{}{\boldsymbol{\lambda}}{\<}\Omega|\chi_{\alpha} (\mathbf{x}, \tau)\chi^{\dagger}_{\beta}(0)|\Omega\>_{\boldsymbol{\lambda}},
             \label{pertprop}
     \end{split}
\end{equation}
where $\boldsymbol{\lambda}=(\lambda_1, \lambda_2)$, and $\Gamma$ is the spin-parity projector. 

\begin{widetext}
Inserting two complete sets of states and taking $\chi (\tau) = e^{-H_{\text{FH}}\tau}\chi (0)$, Eq.~\eqref{pertprop} becomes
\begin{equation}
    \begin{split}
      &  \mathcal{G}_{\boldsymbol{\lambda}} (\tau, \mathbf{p}') = \frac{1}{4}\sum_{s,s'}\Gamma_{\beta\alpha}\sum_{X,Y}\int \frac{d^3p}{(2\pi)^3}\frac{ \prescript{}{\boldsymbol{\lambda}}{\<}\Omega|\chi (0)_{\alpha}|X(\mathbf{p}',s')\>\<X(\mathbf{p}',s')|e^{-H_{\text{FH}}\tau}|Y(\mathbf{p},s)\>
      \<Y(\mathbf{p},s)|\chi^{\dagger}_{\beta}(0)|\Omega\>_{\boldsymbol{\lambda}}}{4E_X(\mathbf{p}')E_Y(\mathbf{p})}.
      \label{pertprop2}
     \end{split}
\end{equation}
Note that states and energies without a $\lambda$ subscript are unperturbed.

We can expand the time evolution operator, $e^{-H_{\text{FH}}\tau}$, with a Dyson series:
\begin{equation*}
    e^{-H_{\text{FH}}\tau} = e^{-H_{\text{QCD}}\tau}\bigg [1 +\sum_{j=1,2}\lambda_j\int _{0}^{\tau}d\tau_1 V_j(\tau_1) + \sum_{j,k=1,2}\lambda_j\lambda_k \int _{0} ^{\tau}d\tau _1 \int _{0}^{\tau_1} d\tau_2 V_j(\tau_1)V_k(\tau_2) \bigg ]+\mathcal{O}(\lambda^3),
\end{equation*}
and hence Eq.~\eqref{pertprop2} becomes
\begin{equation}
    \begin{split}
       &  \mathcal{G}_{\boldsymbol{\lambda}} (\tau, \mathbf{p}')
        = \sum_{X,Y}\int \frac{d^3p}{(2\pi)^3}\frac{1}{4E_X(\mathbf{p}')E_Y(\mathbf{p})}\prescript{}{\boldsymbol{\lambda}}{\<}\Omega|\chi (0)|X(\mathbf{p}')\>\<Y(\mathbf{p})|\chi^{\dagger}(0)|\Omega\>_{\boldsymbol{\lambda}}
     \\ &  \times \<X(\mathbf{p}')|e^{-H_{\text{QCD}}\tau}\bigg [1 +\sum_{j=1,2}\lambda_j\int _{0}^{\tau}d\tau_1 V_j(\tau_1) + \sum_{j,k=1,2}\lambda_j\lambda_k \int _{0} ^{\tau}d\tau _1 \int _{0}^{\tau_1} d\tau_2 V_j(\tau_1)V_k(\tau_2) +\mathcal{O}(\lambda^3)\bigg ]|Y(\mathbf{p})\>.
        \label{pertpropexpansion}
    \end{split}
\end{equation}

\begin{center}
    \begin{figure}
    \centering
    \includegraphics[scale=0.85]{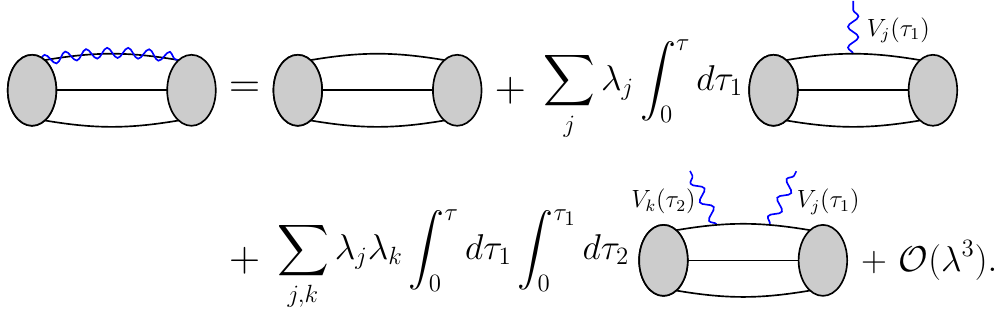}
    \caption{Illustration of the expansion of the perturbed propagator, Eq.~\eqref{pertpropexpansion}, where we have suppressed the tower of states at source and sink. Euclidean time increases left to right.}
    \label{fig:pertpropfig}
\end{figure}
\end{center}
\end{widetext}
Note that we have dropped the spin structure for brevity, but will reintroduce it in the final result.

From Eq.~\eqref{pertpropexpansion}, we see that the $\mathcal{O}(\lambda^2)$ terms of the two-point propagator contain four-point functions (see Figure \ref{fig:pertpropfig}). In particular, the $\lambda_i^2$ term has both currents inserting momentum $\pm\mathbf{q}_i$, and hence provides access to the \emph{forward} Compton amplitude. Only the mixed, second-order term, proportional to $\lambda_1\lambda_2$, will have different incoming/outgoing momenta, and therefore off-forward kinematics.

To isolate the mixed second-order term, we define the combination of nucleon propagators,
\begin{equation}
    R_{\lambda} \equiv \frac{\mathcal{G}_{(\lambda,\lambda)}+\mathcal{G}_{(-\lambda,-\lambda)}-\mathcal{G}_{(\lambda,-\lambda)}-\mathcal{G}_{(-\lambda,\lambda)}}{\mathcal{G}_{(0,0)}}.
    \label{combocorr}
\end{equation}
Having established how to isolate the second-order, off-forward contribution to the perturbed propagator, we are now interested in how to ensure ground state saturation at the source and sink.

As detailed in Ref.~\cite{fwdpaper}, provided that none of the intermediate states are lower energy than $E_N$, we have
\begin{equation*}
  \sum_X  \<X(\mathbf{p}')|e^{-H_{\text{QCD}}\tau} \overset{\tau\gg a}{\simeq}\<N(\mathbf{p}')|e^{-E_N\tau}.
\end{equation*}
%Note that we will use $N$ to refer to the ground state, even though our discussion generalises to hadrons that are not nucleons.
Using this result, Eq.~\ref{combocorr} becomes
\begin{equation}
    \begin{split}
       &  R_{\lambda} (\tau, \mathbf{p}')
        \overset{\tau\gg a}{\simeq} 4\lambda^2\sum_{Y}\int \frac{d^3p}{(2\pi)^3}\frac{A_Y^{\lambda}(\mathbf{p}')}{2E_Y(\mathbf{p})}
        \\ &\times \bigg[
     \int _{0} ^{\tau}d\tau _1 \int _{0}^{\tau_1} d\tau_2 \<N(\mathbf{p}')|V_1(\tau_1)V_2(\tau_2)  |Y(\mathbf{p})\>
     \\ & +\int _{0} ^{\tau}d\tau _1 \int _{0}^{\tau_1} d\tau_2 \<N(\mathbf{p}')|V_2(\tau_1)V_1(\tau_2)  |Y(\mathbf{p})\>\bigg],
        \label{propcomboexpansion}
    \end{split}
\end{equation}
neglecting $\mathcal{O}(\lambda^4)$ corrections and where
$$ A^{\lambda}_Y(\mathbf{p}')= \frac{{\<}\Omega|\chi (0)|N(\mathbf{p}')\>\<Y(\mathbf{p})|\chi^{\dagger}(0)|\Omega\>}{|\<N(\mathbf{p}')|\chi(0)|\Omega\>|^2} + \mathcal{O}(\lambda^2).
$$

Unlike a direct four-point function approach, ground state saturation at the source is ensured by a judicious choice of kinematics, not by large Euclidean time separations---see appendix \ref{sec:appendixFH} for a full calculation. To summarise, these kinematic restrictions require that the current insertion momenta, $\mathbf{q}_1$ and $\mathbf{q}_2$, and sink momentum, $\mathbf{p}'$, are chosen such that 
\begin{itemize}
    \item $|\mathbf{p}'|\leq|\mathbf{p}+n\mathbf{q}_1+m\mathbf{q}_2|$ for $m,n\in \mathbb{Z}$, which prevents the intermediate states from going on-shell,
    \item and $|\mathbf{p}'| = |\mathbf{p}'+\mathbf{q}_1-\mathbf{q}_2|$, which keeps the incoming and outgoing states energy degenerate.
\end{itemize} 

After these restrictions are imposed, Eq.~\eqref{propcomboexpansion} can be written, up to $\mathcal{O}(\lambda^4)$ corrections, as
\begin{equation}
    \begin{split}
        & R_{\lambda} (\tau, \mathbf{p}')
        \overset{\tau\gg a}{\simeq} \lambda^2 \mathcal{C}+\frac{2\lambda^2}{E_N(\mathbf{p}')}
       \\ & \times\frac{\sum _{s,s'} \text{tr}\big[\Gamma u(\mathbf{p}',s')T^{33}(\mathbf{p}', \mathbf{q}_1, \mathbf{q}_2; s',s)  \bar{u}(\mathbf{p},s)\big]}{\sum _{s}{\text{tr}[\Gamma u(\mathbf{p}',s)\bar{u}(\mathbf{p}',s)]}},
        \label{spinpropcomboexpansion}
    \end{split}
\end{equation}
where $T^{33}$ is the $\mu=\nu=3$ component of the OFCA for a single quark flavour with unit charge, and $\mathbf{p}=\mathbf{p}'+\mathbf{q}_1-\mathbf{q}_2$. The term $\mathcal{C}$ is constant in both $\lambda$ and $\tau$; it is made up of contributions for which the source is not the ground state (see Eq.~\eqref{intercept}). %The terms that have ground state saturation at the source are enhanced by a factor of the Euclidean time, $\tau$, with a coefficient proportional to the off-forward Compton amplitude. 

Therefore, by fitting $R_{\lambda}(\tau, \mathbf{p})$ in $\tau$ and $\lambda$, we can isolate the OFCA.

\section{The Off-Forward Compton Amplitude} \label{sec:gca}

Given the method to calculate the OFCA presented in the previous section, we now show how to parameterise the invariant amplitudes of the tensor decomposition, Eq.~\eqref{expltensordecomp}, in terms of GPDs. 

%Almost all existing twist expansions of the OFCA (see a classic example in Ref.~\cite{jiog}) restrict themselves to a frame in which the kinematics are dominated by collinear light-like vectors. As such, the wealth of existing literature on twist expansions of the OFCA can not be directly compared to a lattice calculation. 

The suitable tool for a perturbative expansion of the OFCA in the Euclidean region is the operator product expansion (OPE), which is an expansion about points in coordinate space and momentum space ($z^{\mu}= 0$ and $\bomega=0$, respectively) that are accessible in a spacetime with Euclidean signature \cite{collinsrenorm}. There exist in the literature several OPEs of the OFCA \cite{watanbe1, watanabe2, chen, white}. However, as these largely focus on the spin-zero case and/or significantly pre-date GPDs, in this section we give our own OPE.

%, as well as expansions using the related light-ray operator formalism \cite{lightray, belitskymuller, radyushkinweiss},
\vspace{4mm}
\subsection*{Operator Product Expansion}
\vspace{-2mm}
 Formally, the leading-twist contribution to the coordinate-space current product is given by the `handbag' contributions \cite{muta, collinsrenorm}:
\begin{align}
\begin{split}
 &  T\{ j_{\mu}(z/2)j_{\nu}(-z/2)\} = -2  \frac{i}{2\pi^2}\frac{z^{\mu}}{(z^2-i\epsilon)^2} 
 \\ & \times\bigg [\mathcal{S}_{\mu\rho\nu\kappa}\sum_{n=1,3,5}^{\infty}\frac{(-i)^n}{n!}z_{\mu_1}...z_{\mu_n}\mathcal{O}^{(n+1)\kappa\mu_1...\mu_{n}}_q(0)
 \\ & + i\varepsilon_{\mu\nu\rho\kappa}\sum_{n=0,2,4}^{\infty}\frac{(-i)^n}{n!}z_{\mu_1}...z_{\mu_n}\tilde{\mathcal{O}}^{(n+1)\kappa\mu_1...\mu_{n}}_q(0) \bigg ],
   \label{currentprod}
\end{split}
\end{align}
where $\mathcal{S}_{\mu\rho\nu\kappa} = g_{\mu\rho}g_{\nu\kappa} +g_{\mu\kappa}g_{\nu\rho} - g_{\mu\nu}g_{\rho\kappa}$, and the operators are defined in Eqs.~\eqref{localtwisttwo} and \eqref{localtwisttwopol}.

To obtain the leading-twist OFCA, Eq.~\eqref{vcadef}, we must take the off-forward matrix element of Eq.~\eqref{currentprod} and Fourier transform it. Details of this calculation are presented in appendix \ref{sec:appendixOPE}.

%Note that this contribution only dominates the current product of the OFCA if $t\ll \bar{Q}^2$ \cite{diehlhandbag}. 

%Therefore, to obtain the leading-twist contribution of the OFCA, we must take the current product in Eq.~\eqref{currentprod} between states $\<P'|$ and $|P\>$. Then, we perform a Fourier transform, as per the OFCA definition, given in Eq.~\eqref{vcadef}. 

The final result is
 \begin{widetext}
 \begin{align}
    \begin{split}
         T^{\{\mu\nu\}}(\bomega, \vartheta, t) & =   \sum_{n=2,4,6}^{\infty}\sum_{j=0,2,4}^{n-1}\bigg \{ \frac{4}{\bar{Q}^2}\frac{1}{n}\bomega^{n-2}(-2\vartheta/\bomega)^{j}[h^{\{\mu}A^q_{n,j}(t)+e^{\{\mu}B^q_{n,j}(t)]
       \Big (\bomega\bar{q}^{\nu\}}+2\bar{P}^{\nu\}}\Big)
       \\ & +\frac{8}{(\bar{Q}^2)^2}\frac{1}{n}\bomega^{n-3}(-2\vartheta/\bomega)^{j}[A^q_{n,j}(t)h\cdot \bar{q}+B^q_{n,j}(t)e\cdot \bar{q}] \Big ((n-1)\bomega\bar{P}^{\{\mu}\bar{q}^{\nu\}} + (n-2)\bar{P}^{\mu}\bar{P}^{\nu} \Big)
          \\ & + \frac{8}{(\bar{Q}^2)^2}\delta_{j,0}\bomega^{n-3}(-2\vartheta/\bomega)^{n}C^q_{n}(t)(h\cdot\bar{q}-e\cdot\bar{q})\Big(\bomega\bar{P}^{\{\mu}\bar{q}^{\nu\}}+\bar{P}^{\mu}\bar{P}^{\nu}\Big)
      \\ & 
      -\frac{2}{\bar{Q}^2}g^{\mu\nu}\bomega^{n-1}\Big((-2\vartheta/\bomega)^j[A^q_{n,j}(t)h\cdot \bar{q}+B^q_{n,j}(t)e\cdot \bar{q}]
      +\delta_{j,0}(-2\vartheta/\bomega)^{n}C^q_{n}(t)(h\cdot\bar{q}-e\cdot\bar{q})\Big)\bigg \},
      \label{yuck}
        \end{split}
        \end{align}
        for the symmetric in $\mu\leftrightarrow \nu$ contribution to the OFCA defined in Eq.~\eqref{vcadef}, while for the anti-symmetric contribution, 
         \begin{equation}
    \begin{split}
         T^{[\mu\nu]}(\bomega, \vartheta, t) = &  \frac{2}{\bar{Q}^2} i\varepsilon^{\mu\nu\rho\kappa}\sum_{n=1,3,5}^{\infty}\sum_{j=0,2,4}^{n-1}\bomega^{n-2}(-2\vartheta/\bomega)^{j}\bigg \{ \frac{1}{n}\Big[\tilde{h}_{\kappa}\tilde{A}^q_{n,j}(t)+\tilde{e}_{\kappa}\tilde{B}^q_{n+1,j}(t)\Big]\bomega\bar{q}_{\rho}
        \\ &  +\frac{2}{\bar{Q}^2}\frac{n-1}{n}\bar{P}_{\kappa}\bar{q}_{\rho}\Big [ \tilde{A}^q_{n,j}(t)\tilde{h}\cdot\bar{q} +\tilde{B}^q_{n+1,j}(t)\tilde{e}\cdot\bar{q}\Big]\bigg \},
      \label{polyuck}
        \end{split}
        \end{equation}
 \end{widetext}
where we have used the bilinear definitions given in Eq.~\eqref{bilineardef}. Recall that the usual skewness variable is $\xi = \vartheta/\bomega$ in our chosen scalars.

 One can verify, by taking the Sudakov decomposition and DVCS kinematics $\bomega \simeq \xi^{-1}$, $\vartheta\simeq1$, that Eqs.~\eqref{yuck} and \eqref{polyuck} recover the standard twist-two DVCS amplitude \cite{jidvcs}. 
 %(Note the factor of 2 difference in the definition of $\xi$ in that paper and the present one.)

  Further, notice that Eqs.~\eqref{yuck} and \eqref{polyuck} violate electromagnetic (EM) gauge invariance (their Ward identities) by terms linear in $\Delta_{\perp}^{\mu} = \Delta^{\mu} +(2\vartheta/\bomega)\bar{P}^{\mu}$. It has been found that the necessary tensor structures to restore EM gauge invariance appear when one considers higher-twist contributions to the handbag diagrams \cite{ belitskymuller, radyushkinweiss2, radyushkinweiss, white}. Therefore, we simply introduce the necessary tensor structures, $\Delta_{[\mu}\bar{P}_{\nu]}$ etc., which restore EM gauge invariance to Eqs.~\eqref{yuck} and \eqref{polyuck}.
  
  %This property of the leading-twist OFCA has been studied in detail in the light-ray operator formalism \cite{ belitskymuller, radyushkinweiss2, radyushkinweiss} and local OPE formalism \cite{white}. It has been found that EM gauge invariance can be restored by considering `total' derivative operators, which are contained in the higher-twist contributions of the handbag diagrams. The process of including the total derivatives in the OPE introduces $\bar{P}^{[\mu}\Delta^{\nu]}$, $\tau_1^{[\mu}\Delta^{\nu]}$, $\tau_2^{[\mu}\Delta^{\nu]}$, and $\Delta^{\mu}\Delta^{\nu}$ terms in the tensor structure of Eqs.~\eqref{yuck} and \eqref{polyuck}, which restore EM gauge invariance. 

We can now use the OPE results, Eqs.~\eqref{yuck} and \eqref{polyuck}, to interpret the high energy limit of each of the scalar amplitudes in the tensor decomposition, Eq.~\eqref{expltensordecomp}:
\begin{itemize}
    \item The scalar amplitudes either vanish at leading-twist, or can be parameterised in terms of convolutions of GPDs. For instance:
    $$  \mathcal{H}_1(\bomega, \vartheta, t) = {2}\sum_{n=2,4,6}^{\infty}\bomega^{n}\int_{-1}^1dxx^{n-1}H(x,\vartheta/\bomega,t).
    $$
    See Eq.~\eqref{loamplitudes} for a full list.
    \item We have off-forward equivalents of the Callan-Gross relation \cite{callangross}: $$\mathcal{H}_1 = \frac{\bomega}{2}\big(\mathcal{H}_2 + \mathcal{H}_3 \big), \quad \mathcal{E}_1 = \frac{\bomega}{2}\mathcal{E}_2.$$ In the forward case, Feynman-Hellmann methods have recently been used to determine power-suppressed Callan-Gross breaking terms \cite{utkucallangross}.
    \item The moments of polarised scalar amplitudes have the following relation at leading-twist
    \begin{equation*}
        \begin{split}
          &  \int ^1 _0 dx x^n \text{Im}\tilde{\mathcal{H}}_1(1/x,\vartheta/\bomega,t) 
          \\ & = -\frac{n+1}{n}\int ^1 _0 dx x^n \text{Im}\tilde{\mathcal{H}}_2(1/x,\vartheta/\bomega,t) ,
        \end{split}
    \end{equation*}
    and similarly for the replacement $\tilde{\mathcal{H}}\to\tilde{\mathcal{E}}$. In the forward limit, this reduces to a relation between the spin-dependent structure functions \cite{spindependenttwist2}.
    \item The $\mathcal{K}$ scalar amplitudes vanish at leading-twist, but do contribute at twist-three in terms of transverse GPDs \cite{diehltransverse, belitskymullerkirchner, dv2quasireal}.
    \item The leading-twist contribution to the subtraction function, Eq.~\eqref{subDR}, is $$S_1(\vartheta, t) = 2 \sum_{n=2,4,6}^{\infty}(2\vartheta)^nC_n(t),$$ which has been studied in relation to the $D$-term \cite{teryaev, anikinteryaev, diehlivanov, pasquini}.
\end{itemize}

\subsection*{Parameterisation of the lattice calculation}

From the Feynman-Hellmann relation, Eq.~\eqref{spinpropcomboexpansion}, we calculate
\begin{equation}
     \frac{\sum _{s,s'} \text{tr}\big[\Gamma u(P',s')T^{33}  \bar{u}(P,s)\big]}{\sum _{s}{\text{tr}[\Gamma u(P',s)\bar{u}(P',s)]}} \equiv \mathcal{R}(\bomega, t,\bar{Q}^2).
    \label{lattcompton}
\end{equation}
Therefore, to get a parameterisation that can be compared to the lattice, we use the tensor decomposition of section \ref{sec:bckgrnd} and the OPE with the following additional conditions:
\begin{itemize}
    \item We choose the $\mu=\nu=3$ component of our Compton amplitude.
    \item The Feynman-Hellmann relation requires $\bar{q}^4=\Delta^4=0$.    
    \item We use zero-skewness ($\xi=0 = \vartheta$) kinematics by choosing $\mathbf{q}_1^2=\mathbf{q}_2^2$.
    \item We use the spin-parity projector $\Gamma = (\mathbb{I}+\gamma_4)/2$.
\end{itemize}
The zero-skewness condition removes the tensor structures with scalar amplitudes $\mathcal{K}_{3,4, 6, 8,9}$ and $\tilde{\mathcal{E}}_{2}$. Further, by calculating the $\mu=\nu=3$ component and taking a spin trace, the tensor structures associated with the polarised amplitudes $\tilde{\mathcal{H}}$ and $\tilde{\mathcal{E}}$ are made irrelevant.

%Further, by calculating the $\mu=\nu=3$ component and taking a spin trace, we remove tensor structures associated with $\tilde{\mathcal{H}}_1$ and $\tilde{\mathcal{E}}_1$. While there is a contribution from the tensor structure with amplitude $\tilde{\mathcal{H}}_2$, it is suppressed by a factor of $1/\bar{Q}^4$. Hence it is very small compared to the $\mathcal{H}$ and $\mathcal{E}$ terms.

Finally, since we take $\bar{Q}^2\sim 7\;\text{GeV}^2$, we will consider the remaining amplitudes, $\mathcal{K}_{1,2,5,7}$ to be suppressed, since they have no leading-twist contribution.

Although a more complete study of the $\bar{Q}^2$-dependence is essential, for this exploratory work we will neglect the $\bar{Q}^2$ suppressed $\mathcal{K}_{1,2,5,7}$ amplitudes, keeping only the $\mathcal{H}_{1,2,3}$ and $\mathcal{E}_{1,2}$ amplitudes.

Therefore, Eq.~\eqref{lattcompton} is
\begin{widetext}
\begin{align}
    \begin{split}
          & \mathcal{R}(\bomega, t,\bar{Q}^2)  =
           \frac{1}{E_N+m_N}\bigg \{\delta_{\rho\sigma}\big[ (E_N+m_N)\mathcal{H}_1+\frac{t}{4m_N}\mathcal{E}_1\big]
          +\frac{\bar{P}_{\rho}\bar{P}_{\sigma}}{\pq}\Big[(E_N+m_N)\big( \mathcal{H}_2 + \mathcal{H}_3\big) +\frac{t}{4m_N}\mathcal{E}_2\Big] \bigg \}\mathcal{P}_{3\rho}\mathcal{P}_{\sigma 3},
          \label{latticedecomp}
    \end{split}
\end{align}
\end{widetext}
with $\mathcal{P}^{\mu\nu}$ as defined in Eq.~\eqref{gaugeproj}, and using Euclidean conventions now to match the lattice.

Next, we subtract off the $\bomega=0$ contribution:
\begin{equation}
    \begin{split}
         \overline{\mathcal{R}}(\bomega,  t,\bar{Q}^2) =\mathcal{R}(\bomega, t,\bar{Q}^2)- \mathcal{R}(\bomega=0, t,\bar{Q}^2),
    \end{split}
\end{equation}
which is equivalent to replacing $\mathcal{H}_1 \to \overline{\mathcal{H}}_1$ and $\mathcal{E}_1 \to \overline{\mathcal{E}}_1$ in Eq.~\eqref{latticedecomp}.

As in our previous study of the forward Compton amplitude, we find anomalous asymptotic behaviour of the $S_1$ subtraction function. A method for controlling this behaviour has been presented in the forward case, where the anomalous behaviour of $S_1$ is found to have minimal effect on the $\omega$-dependence \cite{interlacingsubtraction}. An extension to the OFCA is a goal of future work.

We then take only the leading-twist contributions to the amplitudes, a full list of which is given in Eq.~\eqref{loamplitudes}. 

Imposing the off-forward Callan-Gross relation reduces the number of linearly independent amplitudes in Eq.~\eqref{latticedecomp} from five to two. The final form is then
\begin{widetext}
\begin{equation}
    \begin{split}
       \overline{\mathcal{R}}(\bomega,  t,\bar{Q}^2) = 2K_{33} \sum_{n=2,4,6}^{\infty}\bomega ^{n}\big[A^q_{n,0}(t)+\frac{t}{4m_N(E_N+m_N)}B_{n,0}^q(t)\big],
         \label{nucleonLOFH}
    \end{split}
\end{equation}
\end{widetext}
where $E_N = \sqrt{m_N^2 + \mathbf{p}^2}$ is the sink energy, and
\begin{equation}
    K_{\mu\nu} = \frac{\bar{P}_{\mu}\bar{q}_{\nu}+\bar{P}_{\nu}\bar{q}_{\mu} + \Delta_{[\mu}\bar{P}_{\nu]}}{\bar{P}\cdot \bar{q}}+\frac{\bar{Q}^2}{(\bar{P}\cdot \bar{q})^2}\bar{P}_{\mu}\bar{P}_{\nu}+\delta_{\mu\nu}.
    \label{kinfac}
\end{equation}
For a first approximation of extracting the GPD moments, we will calculate $$\overline{\mathcal{R}}(\bomega, t, \bar{Q}^2)/K_{33}(\bar{P}_3, \bar{q}_3, \pq, \bar{Q}^2).$$

Since our lattice calculations are in frames that are roughly near the rest frame (i.e.~$E_N \approx m_N$), we can approximately treat the combination of GFFs in Eq.~\eqref{nucleonLOFH} as a Lorentz scalar:
\begin{equation}
    \begin{split}
    M^q_{n}(t) & \equiv  A^q_{n,0}(t)
   +\frac{t}{8m_N^2}B_{n,0}^q(t).
  \label{momdefinition}
    \end{split}
\end{equation}
A determination of the $A$ and $B$ GFFs independently, rather than the linear combination defined in Eq.~\eqref{momdefinition}, is desirable. To this end, note that we can also use the spin-parity projector,
$$ \Gamma = \frac{1}{2}(\mathbb{I}+\gamma_4)\gamma_k\gamma_5, \quad k=1,2,3,
$$
which would give linearly independent combinations of the $A$ and $B$ form factors compared with Eq.~\eqref{nucleonLOFH}, in a manner analogous to the separation of $F_1$ and $F_2$ electromagnetic form factors. Hence a separation of the $A$ and $B$ form factors by varying the spin-parity projector is a goal of future work.

\section{Simulation Details} \label{sec:numcalc}

For this calculation, we use the same gauge ensembles as Ref.~\cite{fwdpaper}. Note, in particular, that we are at the SU(3) flavour symmetric point, $\kappa_l=\kappa_s$, with a larger-than-physical pion mass, $m_{\pi}=466(13)\;\text{MeV}$, and a lattice spacing of $a=0.074(2)\;\text{fm}$. See Table \ref{tab:gauge_details} for a summary of the gauge configurations. 
		\begin{table*}
				\centering
				\caption{ \label{tab:gauge_details} Details of the gauge ensembles used in this work.}
				\setlength{\extrarowheight}{2pt}
	    	\begin{tabularx}{\textwidth}{CCCCCCCCCCC}
					\hline\hline
					$N_f$ & $c_{SW}$ & $\kappa_l$ & $\kappa_s$ & $L^3 \times T$ & $a$ & $m_\pi$ & $m_N$ & $m_\pi L$ & $Z_V$ & $N_\text{cfg}$\\
					&&&&& [fm] &[GeV]&[GeV]&&& \\
					\hline
					$2+1$ & 2.65 & 0.1209 & 0.1209 & $32^3\times64$ & 0.074(2) & $0.467(12)$ & $1.250(39)$ & $\sim 5.6$ & 0.8611(84) & 1763 \\
					\hline\hline
				\end{tabularx}
			\end{table*}

\subsection*{Feynman-Hellmann Implementation}

The Feynman-Hellmann implementation is almost identical to our previous study of the forward Compton amplitude \cite{fwdpaper}. In practice, the objects we calculate are perturbed quark propagators, given by
\begin{equation}
   S_{\boldsymbol{\lambda}}(x_n - x_m)= \big [M - \lambda_1\mathcal{O}_1- \lambda_2\mathcal{O}_2 \big]^{-1}_{n,m},
\end{equation}
where $M$ is the usual fermion matrix.

For our case, where we choose to calculate the $\mu=\nu=3$ component of the OFCA, the operators are
$$  [\mathcal{O}_k]_{n,m} = \delta_{n,m}(e^{i\mathbf{q}_k\cdot \mathbf{n}}+e^{-i\mathbf{q}_k\cdot \mathbf{n}})i\gamma_3, \;\; k=1,2.
$$

%Our lattice action is
%\begin{equation}
%    \begin{split}
%      &  S = S_{\text{QCD}} +\int d^3 z \Big( \lambda_1(e^{i\mathbf{q}_1\cdot\mathbf{z}}+e^{-i\mathbf{q}_1\cdot\mathbf{z}})J(z)
%        \\ & + \lambda_2(e^{i\mathbf{q}_2\cdot\mathbf{z}}+e^{-i\mathbf{q}_2\cdot\mathbf{z}}) J(z)\Big).
%    \end{split}
%\end{equation}
Then, the usual formulae for hadrons in terms of quark propagators apply, except with one or more of these propagators replaced with a perturbed propagator.

The Feynman-Hellmann perturbation is applied to the connected contributions only. While it is possible to perturb the disconnected contributions, this would be much more computationally expensive \cite{nonpertrenorm, collabss}.

The determination of the ratio in Eq.~\eqref{combocorr} requires four separate sets of correlators at each magnitude of $\lambda$. We calculate two magnitudes of $\lambda = 0.0125, 0.025$, chosen based on $\lambda$-tuning tests carried out in the forward case \cite{kimthesis, fwdpaper}

\subsection*{Kinematics}

We calculate two sets of correlators on the same gauge configurations (Table \ref{tab:q_kinematics}).
%\begin{itemize}
%    \item Set \#1: $\mathbf{q}_1 = \frac{2\pi}{L}(1,5,1)$ and $\mathbf{q}_1 = \frac{2\pi}{L}(-1,5,1)$, which gives $t=-1.10\;\text{GeV}^2$, $\bar{Q}^2=7.13\;\text{GeV}^2$.
%    \item Set \#2: $\mathbf{q}_1 = \frac{2\pi}{L}(4,2,2)$ and $\mathbf{q}_1 = \frac{2\pi}{L}(2,4,2)$, which gives $t=-2.20\;\text{GeV}^2$, $\bar{Q}^2=6.03\;\text{GeV}^2$. 
%\end{itemize}

\begin{table}[h!]
    \centering
    \caption{Current insertion momenta, $\mathbf{q}_{1,2}$, and derived kinematics for two sets of correlators.}
    \setlength{\extrarowheight}{5pt}
					{
			\begin{tabularx}{.485\textwidth}{C|C|C|C|C}
				\hline\hline
			 	Set & 	$\frac{L}{2\pi}\mathbf{q}_1$,	$\frac{L}{2\pi}\mathbf{q}_2$ & $t\;[\text{GeV}^2]$ &  $\bar{Q}^2\;[\text{GeV}^2]$ &$N_{\text{meas}}$ 	\\
				\hline
				\multirowcell{1}{\#1} 
			                        &	$(1,5,1)$ $(-1,5,1)$ & $-1.10$ & $7.13$ & $996$ \\ 
				\hline
				\multirowcell{1}{\#2} 
			                        &	 $(4,2,2)$ $(2,4,2)$ & $-2.20$ & $6.03$ & $996$ \\ 
				\hline
				\hline
			\end{tabularx}
			}
    \label{tab:q_kinematics}
\end{table}

To fit GPD moments, we need multiple $\bomega$ values. However, we are restricted by the conditions of the Feynman-Hellmann relation to a frame for which our sink momentum, $\mathbf{p}'$, and our momenta from the current insertions, $\mathbf{q}_1$ and $\mathbf{q}_2$, must obey:
$$ |\mathbf{p}'| = |\mathbf{p}'\pm\mathbf{q}_1 \mp \mathbf{q}_2|,
$$
which limits the number of $\bomega$ values that are accessible for each $\mathbf{q}_{1,2}$ pair. 

For each set of correlators, the $\bomega$ value is determined by the value of the sink momentum, $\mathbf{p}'$:
$$ \bomega = \frac{4\mathbf{p}'\cdot (\mathbf{q}_1+\mathbf{q}_2)}{(\mathbf{q}_1+\mathbf{q}_2)^2}.
$$
The explicit values of $\bomega$ for our kinematics are shown in Table \ref{tab:omega_kinematics}.

Moreover, since our amplitude is invariant under the exchanges $\Delta^{\mu}\to-\Delta^{\mu}$, $\bomega \to-\bomega$, we average over $\pm \mathbf{p}'$, $\pm (\mathbf{p}'-\mathbf{q}_1+\mathbf{q}_2)$ to increase our statistics.

\begin{table}[h!]
    \centering
     \caption{$\bomega$ values for the two sets of correlators. Note that $|\bomega|>1$ values are omitted.}
\setlength{\extrarowheight}{5pt}
					{
			\begin{tabularx}{.485\textwidth}{CCC}
				\hline\hline
			Correlator set 	& $\frac{L}{2\pi}\mathbf{p}'$ & $\bomega$	\\
				\hline
				\multirowcell{7}{\#1\\ $t=-1.10\;\text{GeV}^2$
				\\ $\bar{Q}^2=7.13\;\text{GeV}^2$} 
			                        &	 $(1,0,0)$ & 0 \\ 
                                    &  $(1,0,1)$ & $1/13$ \\
                                    &  $(1, 0, 2)$ & $2/13$ \\
                                    &   $(1,1,-1)$ & $  4/13$ \\ 
                                    &   $(1,1,0)$ & $ 5/13$ \\ 
                                    &  $(1,1,1)$ & $ 6/13$ \\
                                    &  $(1,  2,0)$ & $10/13$ \\											
				\hline
					\multirowcell{5}{\#2\\ $t=-2.20\;\text{GeV}^2$
				\\ $\bar{Q}^2=6.03\;\text{GeV}^2$}  
			              &   $(1,-1,0)$ & 0 \\ 
                          &    $(1,-1, 1)$ & $ 2/11$\\ 
                          &    $ (2,0,- 1)$  & $ 4/11$\\ 
                         &     $( 2,0,0)$ & $ 6/11$\\ 
                         &     $ (2,0,1)$ & $ 8/11$\\
				\hline\hline
			\end{tabularx}
			}
    \label{tab:omega_kinematics}
\end{table}

%	\begin{table*}
%				\centering
%				\caption{ \label{tab:FH_momenta} Feynman-Hellmann parameters and momentum insertions for our two sets of correlators.}
%				\setlength{\extrarowheight}{2pt}
%	    	\begin{tabularx}{\textwidth}{C | CC |CC | C | C}
%	    	\hline\hline 
%	    	\# & $\frac{L}{2\pi}\mathbf{q}_1$ & $\frac{L}{2\pi}\mathbf{q}_2$ & $t\;[\text{GeV}^2]$ & $\bar{Q}^2\;[\text{GeV}^2]$  & $\lambda$ & $N_{\text{meas}}$
%				\\ \hline	1	& $(1,5,1)$ & $(-1,5,1)$  & $-1.10$ & $7.13$  & 0.0125, 0.025 & 996 
%				\\ 2	& $(4,2,2)$ & $(2,4,2)$  & $-2.20$ & $6.03$   & 0.0125, 0.025 & 996
%				\\					\hline\hline
%				
%				\end{tabularx}
%			\end{table*}

\section{Results and Discussion} \label{sec:results}

To demonstrate what can be accomplished with the method outlined in the preceding sections, we determine the first two even moments of the nucleon GPD.

\begin{figure}
    \centering
    \includegraphics[width=\linewidth]{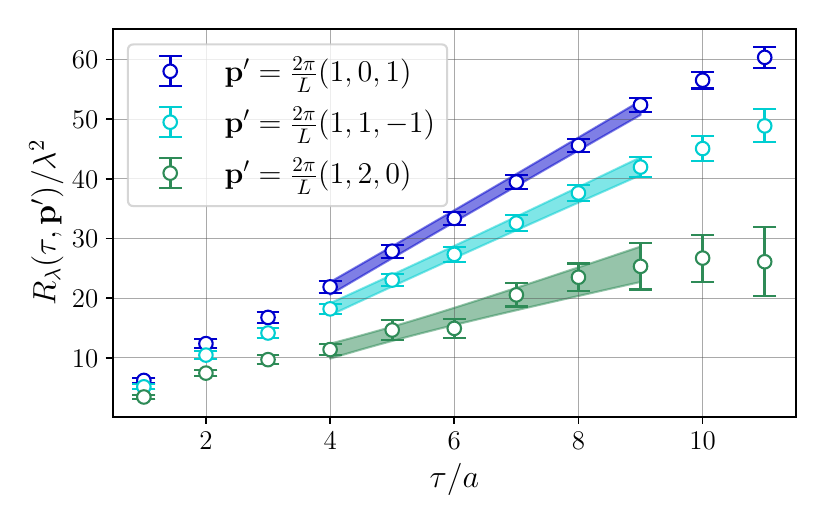}
    \caption{Plot of $\tau$-dependence of $R_{\lambda}/\lambda^2$, as defined in Eq.~\eqref{combocorr}. The shaded bands are fits to the function $f(\tau) = a\tau +b$. The two $\lambda$ magnitudes have been averaged over.}
    \label{taudep}
\end{figure}

First, we fit the combination of correlators, ${R}_{\lambda}(\tau, \mathbf{p}')$ from Eq.~\eqref{combocorr} to the function $f(\tau)= c_1\tau +c_2$, where $\tau$ is Euclidean time. From the Feynman-Hellmann relation, Eq.~\eqref{spinpropcomboexpansion}, the slope, $c_1$, is proportional to the OFCA, while $c_2$ is a superfluous parameter. In fitting this linear function, we apply a consistent fit window in Euclidean time for all sink momenta. The $\chi^2/\text{dof}$ for these fits are reported in Table \ref{tab:fitpar} and it is found that $\chi^2/\text{dof}\sim1$ for all the momenta, which demonstrates that the data is largely well described by a linear fit. An example of the Euclidean time fits for set \#1 is given in Figure \ref{taudep}.

\begin{figure}
    \centering
    \includegraphics[width=\linewidth]{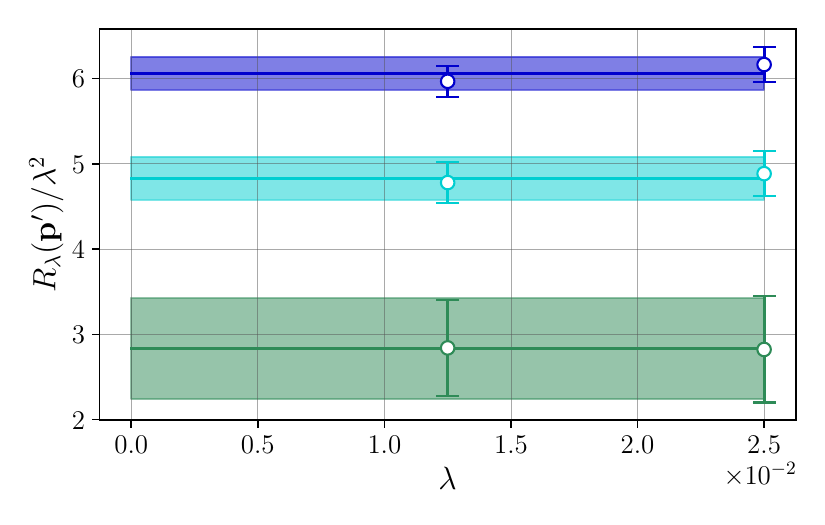}
    \caption{Plot of $\lambda$-dependence of the combination of correlators, $R_{\lambda}/\lambda^2$, as defined in Eq.~\eqref{combocorr}, after fitting in Euclidean time. The shaded bands are the quadratic coefficient for each momentum. Momenta correspond to those in Figure \ref{taudep}.}
    \label{lamdep}
\end{figure}

After the fits in Euclidean time have been performed, we next investigate the behaviour of the ratio, $R_{\lambda}(\mathbf{p}')$, as a function of the Feynman-Hellmann coupling, $\lambda$. From the Feynman-Hellmann relation, Eq.~\eqref{spinpropcomboexpansion}, the $\lambda^2$ contribution to this ratio is proportional to the OFCA, and the next-to-leading contribution is $\mathcal{O}(\lambda^4)$, which is suppressed for our calculations at $\lambda \sim 10^{-2}$.

Therefore, to test the effects of the $\lambda^4$ contributions, we compare the quadratic coefficient of the ratio as extracted with a purely quadratic fit function, $f(\lambda) = b\lambda^2$, to that extracted with the function $g(\lambda) = b\lambda^2+c\lambda^4$. We find that the quartic coefficient, $c$, is consistent with zero, and that the quadratic coefficients, $b$, calculated using the two fit functions agree within errors.

However, since the quartic fit determines two parameters from two $\lambda$ values, it is not a reliable estimate of the higher order contaminations. Therefore, to further examine the effect of these contaminations, we calculate the quotient $(\lambda_1^2R_{\lambda_2})/(\lambda_2^2R_{\lambda_1})$, which is 1 for perfectly quadratic results. In Table \ref{tab:fitpar}, we can see that, although the central value of this quotient is close to 1 for all momenta, not all are within errors of 1. This indicates a $2-4\%$ contamination from higher order terms, which is negligible compared to our overall errors. %Controlling such suppressed, higher order contributions is a current area of investigation \cite{mischarogerdyson}.

Hence for this preliminary study, we find it sufficient to use the purely quadratic fit function, $f(\lambda)=b\lambda^2$. In Figure \ref{lamdep}, we plot the normalised ratio, $R_{\lambda}/\lambda^2$, as a function of $\lambda$, and compare this to the quadratic coefficient from the fit. We observe that the data is reasonably well described by a purely quadratic fit. 

\begin{figure}[t!]
    \centering
    \includegraphics[width=\linewidth]{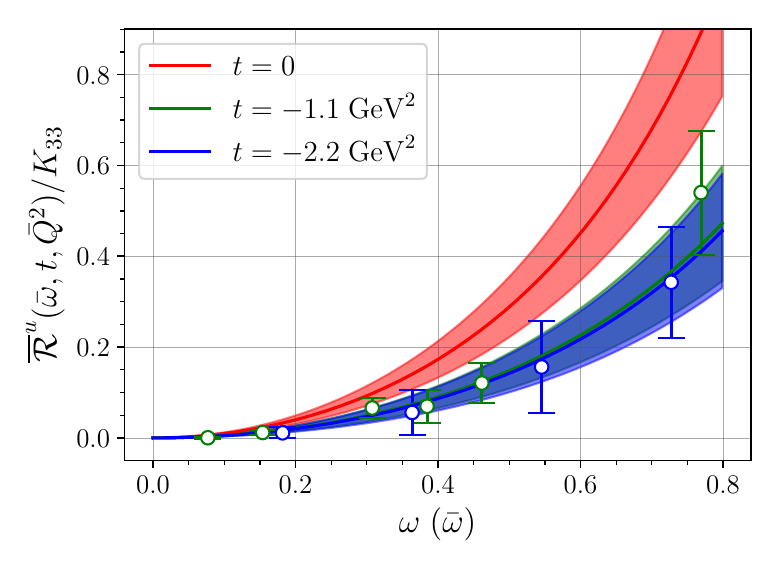}
    \caption{Plot of $\overline{\mathcal{R}}$, as defined in Eq.~\eqref{nucleonLOFH}, divided by the kinematic factor, $K_{33}$, from Eq.~\eqref{kinfac}. The red curve is a parameterisation of results from Ref.~\cite{fwdpaper}. The blue and green curves are from the moment fits.}
    \label{off_fwd_vs_fwd}
\end{figure}

\begin{table}[h!]
    \centering
    \caption{Parameters demonstrating the quality of fits in Euclidean time and the Feynman-Hellmann parameter, $\lambda$, for the up quark results from set \# 1.}
    \setlength{\extrarowheight}{5pt}
					{
			\begin{tabularx}{.485\textwidth}{CCC}
				\hline\hline
			 	$\frac{L}{2\pi}\mathbf{p}'$ & $\chi^2/$dof ($\tau$ fits) & $(\lambda_1^2R_{\lambda_2})/(\lambda_2^2R_{\lambda_1})$ \\ \hline	
			                     (1,0,0)   & 0.87 & 1.039(4) \\ 
			                   (1,0,1)     &	1.1  & 1.033(5) \\ 
			                     (1,0,2)        & 0.75 & 1.01(2) \\ 
			                    (1,1,-1)    &	0.49  & 1.019(6) \\ 
			                      (1,1,0)       & 1.0 & 1.032(4) \\ 
			                    (1,1,1)    &	0.57  & 1.022(6)\\ 
			                  (1,2,0)   &	1.6  & 0.99(3) \\ 
				\hline
				\hline
			\end{tabularx}
			}
    \label{tab:fitpar}
\end{table}

Using the Feynman-Hellmann relation, Eq.~\eqref{spinpropcomboexpansion}, we can now interpret the quadratic coefficient as the off-forward Compton amplitude. Then, by varying the sink momentum, we can calculate the amplitude at multiple values of the scaling variable, $\bomega$. The results for the up quark in the nucleon are shown in Figure \ref{off_fwd_vs_fwd}. 

The forward $t=0$ curve in this plot is a fit to the $Q^2 = 7.13\;\text{GeV}^2$ results from Ref.~\cite{fwdpaper}. As that study also used the Feynman-Hellmann method and the same gauge configurations as the present calculation, we can compare it to our off-forward, $t\neq0$, results to determine the $t$-dependence of the OFCA.

\subsection*{Moment Fitting}

Using the results of our OPE in section \ref{sec:gca}, we can interpret the moments of the OFCA as GPD moments, defined in Eq.~\eqref{momdefinition}. Hence, using Eq.~\eqref{nucleonLOFH}, a fit in $\bomega$ to the function
\begin{equation}
    \begin{split}
        & f_{J}(\bomega, t, \bar{Q}^2) =
        2\sum_{n=2,4,6}^{2J}\bomega ^{n} M_{n}(t, \bar{Q}^2)
        \label{momparam}
    \end{split}
\end{equation}
yields the first $J$ even GPD moments at fixed $t$ and $\bar{Q}^2$ values. At leading-twist, these moments are $$M_n(t) = A_{n,0}(t) + \frac{t}{8m_N^2}B_{n,0}(t). $$

\begin{figure}[t!]
    \centering
    \hspace{-1.0em}
   \includegraphics[width=\linewidth]{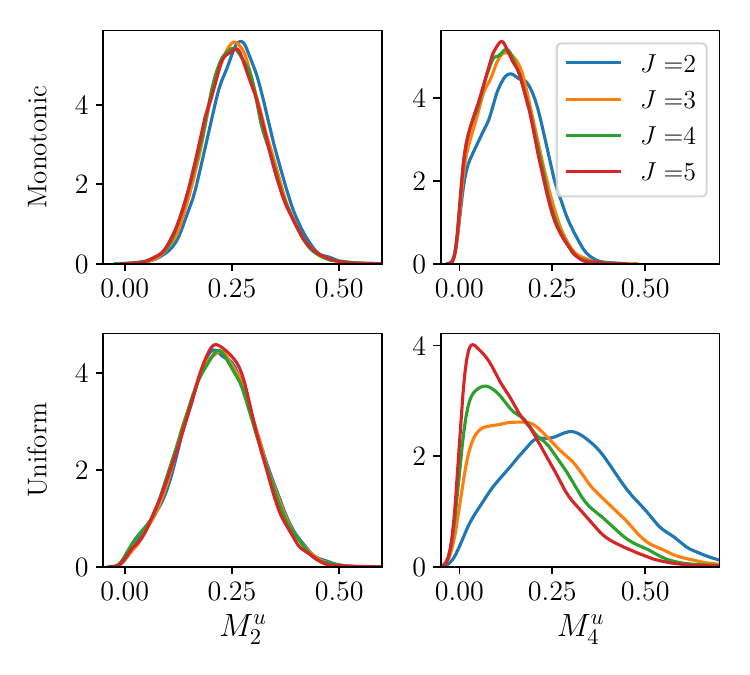}
    \caption{Density distributions for the first two up quark moments at $t=-1.10\;\text{GeV}^2$. The upper two plots use monotonic priors distributions, while the lower two plots use uniform positive distributions. $J$ is the number of moments fit in the parameterisation, Eq.~\eqref{momparam}.}
    \label{momdensityplot}
\end{figure}

Unlike the forward case, there is no optical theorem connecting the OFCA to the scattering cross section, and therefore no requirement for the scalar amplitudes to be positive definite. However, our moments, defined in Eq.~\eqref{momdefinition}, are dominated by $A_{n,0}(t)$, the moments of the zero-skewness GPD $H(x,t)$, which is typically treated as positive in model-dependent parameterisations (for instance, Refs.~\cite{valenceregge1, valenceregge2, valenceregge3, valenceregge4}), while the $E(x,t)$ GPD is suppressed by $t/8m_N^2$ in our moments. Therefore, it is reasonable for this proof-of concept calculation to treat the underlying distribution, $H(x,t) + (t/8m_N^2)E(x,t)$, 
        as strictly positive on the domain $x\in[-1,1]$, and thus its moments as monotonically decreasing for fixed $t$: \begin{equation}
            M_{2}(t)\geq M_{4}(t) \geq ... \geq M_{2J}(t).
            \label{mono_condition}
        \end{equation}

        Future work will aim at a more extensive treatment of the conditions on moments, such as incorporating model-independent positivity constraints on GPDs \cite{Pobylitsa_2002, Pobylitsa_2002_IPS, Pobylitsa_2004} and on the Compton amplitude \cite{compton_positivity}.

To fit these moments, we use a Markov chain Monte Carlo method \cite{pymc3_no1, pymc3_no2}. In contrast to a least squares fit, this method allows us to efficiently sample prior distributions that reflect physically-motivated constraints \cite{fwdpaper}. 

    In Figure \ref{momdensityplot}, we compare the up quark, $t=-1.1\;\text{GeV}^2$ moments fit using monotonically decreasing priors, as in Eq.~\eqref{mono_condition}, to those fit with uniform positive priors, $M_n(t) \in [0,100]$. Since we truncate the series of moments at a finite order, Figure \ref{momdensityplot} also compares the values of the first two moments fit at different orders of truncation, $J$, as in Eq.~\eqref{momparam}. We observe that, for both the monotonic moments, the value of $J$ has little effect on the leading moment, $M_2$. Moreover, the values of $M_2$, as extracted with the monotonic and uniform moments, are highly consistent. 
    
    On the other hand, the value of $M_4$ differs significantly depending on whether uniform or monotonic priors are used. For the uniformly sampled moments, the $M_4$ distributions are heavily skewed towards zero, and do not converge with $J$. By contrast, the monotonically sampled moments, $M_4$, do not depend greatly on the order of truncation for $J>2$, and the distributions appear only slightly skewed towards zero. The higher moments require larger values of $\bomega$ and more precise data to constrain them. Therefore, the inconsistencies in the $M_4$ results likely reflect the fact that we have a limited number of larger $\bomega$ values, which have significant errors. Moreover, these inconsistencies may reflect that the monotonicity condition is too severe for small moments. Investigating these issues is a goal of future studies.

For this preliminary study, we choose to fit the first four even moments, $n=2,4,6,8$, using monotonic conditions, and report the first two even moments. For consistency, we only fit the first four moments of the forward results as well.

\begin{figure}
    \centering
    \hspace{-1.0em}
    \includegraphics[width=\linewidth]{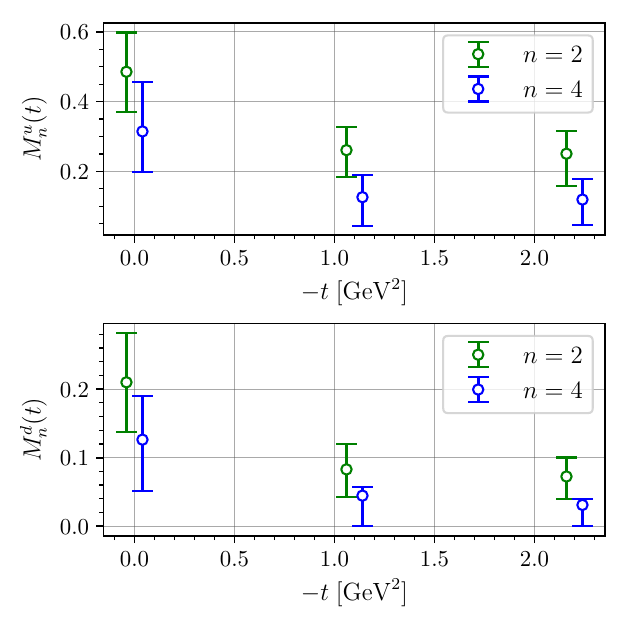}
    \caption{The $t$-dependence of the first two even moments, $M^q_n(t)$, defined in Eq.~\eqref{momdefinition}, for up and down quarks. The $t=0$ points are from a fit to results in Ref.~\cite{fwdpaper}.}
    \label{momplot}
\end{figure}

We present results for the $t$-dependence of the leading moments in Figure \ref{momplot}. The values of the $n=2$ GPD moments are statistically consistent with moments from three-point calculations at a comparable pion mass \cite{gpdlatt6}. However, the $n=4$ moments have never been determined from three-point methods, and therefore the results presented here are a first look at the $t$ behaviour of such moments.

\subsection*{Comment on systematics}

As the present numerical results are exploratory, a detailed assessment of systematic uncertainties remains an objective of future work. A list of the most salient systematics and proposals to control them is given below.

\begin{enumerate}
    \item To better isolate the leading-twist contribution, a range of $\bar{Q}^2$ values must be calculated, and the constant, leading-twist moments fit from this, as in Ref.~\cite{fwdpaper}. 
    \item The two data sets (\#1 and \#2) have different $\bar{q}_3$, which means that the $\mathcal{O}(a)$ Ward identity violating terms, induced by discretisation, will differ between the two data sets. Hence it is preferable to use the conserved vector current, for which exact Ward identities are known \cite{latticeWI1, latticeWI2}.
    \item The OPE performed in section \ref{sec:gca} is a continuum relation, and therefore a continuum extrapolation, similar to that in Refs.~\cite{HOPEpaper, HOPEproceedings, HOPEreport}, is desirable.
    %\item As discussed in the preceding section, the Compton amplitude subtraction function is sensitive to lattice artifacts, which cause unphysical scaling properties. A method for controlling these artifacts has been tested in the forward case, where they were shown have minimal effect on the $\omega$-dependence \cite{interlacingsubtraction}.
    \item Finally, there are all the usual lattice systematics: non-physical quark masses, finite volume, and excited state contamination, which must be accounted for.
\end{enumerate}

\section{Summary and Conclusions} \label{sec:summary}

This study has presented a novel means to determine the off-forward Compton amplitude (OFCA) using lattice QCD, and thereby calculate the properties of generalised parton distributions (GPDs). We derived a Feynman-Hellmann relation to calculate the OFCA. In our parameterisation of the OFCA, we presented new results and collected old ones, which lay the groundwork for comprehensive calculations of GPDs from the OFCA. Finally, the nucleon moments presented here are the first determination of $n=4$ GPD moments.

We are now in a position to realise the full potential of this method. A more detailed investigation of the systematics is a priority, including calculations at different lattice spacings and with the conserved vector current. Currently, such tests are being conducted for the forward Compton amplitude \cite{interlacingsubtraction}. Similarly, future work will be aimed at calculating a greater spread of $\bar{Q}^2$ and $t$, which will provide physical insights and allow us to more accurately determine the leading-twist contribution. Furthermore, we aim to separate out the $\mathcal{H}$ and $\mathcal{E}$ scalar amplitudes---equivalently the $A$ and $B$ generalised form factors. 

Taking these steps would provide us with a wealth of physical information. For instance, we could investigate the non-perturbative features of the OFCA, including the off-forward subtraction function and generalised polarisabilites. Moreover, we could investigate GPD properties, such as their scaling behaviour, and higher-twist contributions to the Compton amplitude. Finally, this method allows us to constrain GPDs, by calculating their moments, fitting models, and other methods to extract parton distributions from the Euclidean Compton amplitude directly \cite{montecarlofit}. 

\section*{Acknowledgements} \label{sec:acknowledgements}

We would like to thank Z.~Kordov for useful discussion and comments on this manuscript. The numerical configuration generation (using the BQCD lattice QCD program~\cite{Haar:2017ubh})) and data analysis (using the Chroma software library~\cite{Edwards:2004sx}) was carried out on the DiRAC Blue Gene Q and Extreme Scaling (EPCC, Edinburgh, UK) and Data Intensive (Cambridge, UK) services, the GCS supercomputers JUQUEEN and JUWELS (NIC, Jülich, Germany) and resources provided by HLRN (The North-German Supercomputer Alliance), the NCI National Facility in Canberra, Australia (supported by the Australian Commonwealth Government) and the Phoenix HPC service (University of Adelaide). AHG is supported by an Australian Government Research Training Program (RTP) Scholarship. RH is supported by STFC through grant ST/P000630/1. PELR is supported in part by the STFC under contract ST/G00062X/1. GS is supported by DFG Grant No. SCHI 179/8-1. KUC, RDY and JMZ are supported by the Australian Research Council grant DP190100297.

\appendix

\onecolumngrid

\section{Background} \label{sec:appendixBG}

For symmetrisation and anti-symmetrisation of a rank-2 tensor, we use the notation
$$ T^{\{\mu\nu\}} = \frac{1}{2}\big[T^{\mu\nu}+T^{\nu\mu}\big], \quad  T^{[\mu\nu]} = \frac{1}{2}\big[T^{\mu\nu}-T^{\nu\mu}\big].
$$
The general expression for a fully symmetrised rank-$n$ tensor used in this paper is
\begin{equation}
    T^{\{\mu_1...\mu_n\}}=\frac{1}{n!}\sum_{\sigma\in S_{n}}T^{\nu_{\sigma(1)}...\nu_{\sigma(n)}},
\end{equation}
where $S_{n}$ is the group of permutations of the numbers $1,2,...,n$, and $\sigma$ is an element of $S_{n}$. Here, we denote the $i^{\text{th}}$ component of some group element, $\sigma\in S_n$, as $\sigma(i)$.

\subsection*{Polarised GPDs}

\begin{itemize}
    \item Polarised light-cone matrix element:
\begin{equation}
\begin{split}
       \int \frac{d\lambda}{2\pi}e^{i\lambda x}\<N(P') & |\bar{\psi_q}(-\lambda n/2)  {\slashed{n}}\gamma_5\psi_q(\lambda n/2)|N(P)\> =\tilde{H}^q(x,\vartheta/\bomega,t)\bar{u}(P')\gamma^{\mu}\gamma_5n_{\mu}u(P)
      \\ &+\tilde{E}^q(x,\vartheta/\bomega,t)\frac{\Delta\cdot n}{2m_N}\bar{u}(P')\gamma_5u(P).
        \label{polGPDdef}
        \end{split}
\end{equation}

\item Local twist-two polarised operators:
\begin{align}
   \begin{split}
    & \tilde{\mathcal{O}}^{(n)\mu_1...\mu_{n}}_q(X) =\bar{\psi}_q(X)\gamma^{\{\mu_1}\gamma_5i{\overset{\leftrightarrow} D}^{\mu_2}... i {\overset{\leftrightarrow} D}^{\mu_n\}} \psi_q(X)- \textnormal{traces}.
    \label{localtwisttwopol}
   \end{split}
\end{align}

\item Their matrix elements:
\begin{equation}
    \begin{split}
        \<N(P')|\tilde{\mathcal{O}}^{(n+1)\kappa\mu_1...\mu_{n}}_q(0) |N(P)\> & = \bar{u}(P',s')\gamma^{\{\kappa}\gamma_5 u(P,s) \sum_{j=0,2,4}^n \tilde{A}^q_{n+1,j}(t)\Delta^{\mu_1}...\Delta^{\mu_j}\bar{P}^{\mu_{j+1}}...\bar{P}^{\mu_n\}}
        \\ & + \frac{\Delta^{\{\kappa}}{2m_N}\bar{u}(P',s')\gamma_5 u(P,s) \sum_{j=0,2,4}^n \tilde{B}^q_{n+1,j}(t)\Delta^{\mu_1}...\Delta^{\mu_j}\bar{P}^{\mu_{j+1}}...\bar{P}^{\mu_n\}}.
        \label{polmatelem}
    \end{split}
\end{equation}

\item Polynomiality:
\begin{equation}
   \begin{split}
       \int _{-1}^1 dx x^{n} & \tilde{H}^q(x,\vartheta/\bomega,t)
      =\sum^n_{i=0,2,4}(2\vartheta/\bomega)^i\tilde{A}^q_{n+1,i}(t),
 \quad \text{and} \quad
  \int _{-1}^1 dx x^{n}  \tilde{E}^q(x,\vartheta/\bomega,t)
 =\sum^n_{i=0,2,4}(2\vartheta/\bomega)^i\tilde{B}^q_{n+1,i}(t).
   \end{split}
   \label{polynomialitypol}
\end{equation}
\end{itemize}

\onecolumngrid

\section{Operator Product Expansion} \label{sec:appendixOPE}

We start with the matrix element of the leading-twist contribution to the current product, Eq.~\eqref{currentprod}. The symmetric under $\mu\leftrightarrow\nu$ component is
\begin{align}
 \begin{split}
 &\<N(P') | T\{ j_{\{\mu}(z/2)j_{\nu\}}(-z/2)\}|N(P)\>  =
 -2  \frac{i}{2\pi^2}\frac{z^{\mu}}{(z^2-i\epsilon)^2} \mathcal{S}_{\mu\rho\nu\kappa}\sum_{n=1,3,5}^{\infty}\frac{(-i)^n}{n!}\sum_{j=0,2,4}^n\Big \{ \frac{1}{n+1}(\Delta\cdot z)^j
 \\ & \times (\bar{P}\cdot z)^{n-j}\big[h^{\kappa}A^q_{n+1,j}(t)+e^{\kappa}B^q_{n+1,j}(t)\big] +\frac{n-j}{n+1} (\Delta\cdot z)^j(\bar{P}\cdot z)^{n-j-1}\bar{P}^{\kappa}\big [ A^q_{n+1,j}(t)h\cdot z+B^q_{n+1,j}(t)e\cdot z \big]
   \\&+\frac{j}{n+1}(\Delta\cdot z)^{j-1}(\bar{P}\cdot z)^{n-j}\Delta^{\kappa}\big [ A^q_{n+1,j}(t)h\cdot z+B^q_{n+1,j}(t)e\cdot z \big]
    + \delta_{j,0}\Delta^{\kappa}(\Delta\cdot z)^nC^q_{n+1}(t)\frac{1}{m_N}\overline{u}(P')u(P)\Big \}.
 \label{matrixelem}
 \end{split}
   \end{align}
The anti-symmetric component is no different to the symmetric component, except with $h (e)\to\tilde{h}(\tilde{e})$, $A^q_{n+1,j}\to\tilde{A}^q_{n+1,j}$, $B^q_{n+1,j}\to\tilde{B}^q_{n+1,j}$, and the $C$ GFFs set to zero.

The general recipe for the Fourier transform of these matrix elements is:

First, introduce Fourier conjugates,
    \begin{align*}
   \begin{split}
       (\bar{P}\cdot z)^n & = i^n\int _{-\infty}^{\infty}d\chi e^{i\chi \bar{P}\cdot z}\frac{\partial^n}{\partial \chi ^n}\delta(\chi),
       \\ (\Delta\cdot z)^n & = i^n\int _{-\infty}^{\infty}d\eta e^{i\eta \Delta\cdot z}\frac{\partial^n}{\partial \eta ^n}\delta(\eta),
       \\  h\cdot z & = i\int _{-\infty}^{\infty}d\tilde{\chi}_1 e^{i\tilde{\chi}_1 h\cdot z}\frac{\partial}{\partial \tilde{\chi}_1 }\delta(\tilde{\chi}_1),
       \\  e\cdot z & = i\int _{-\infty}^{\infty}d\tilde{\chi}_2 e^{i\tilde{\chi}_2 e\cdot z}\frac{\partial}{\partial \tilde{\chi}_2 }\delta(\tilde{\chi}_2).
   \end{split}
\end{align*}
For the polarised component $h (e)\to\tilde{h}(\tilde{e})$, but otherwise the process is the same.

Next, we use the identity
\begin{align*}
\begin{split}
   & \int d^4ze^{il\cdot z}\frac{z^{\mu}}{2\pi^2(z^2-i\epsilon)^2}=\frac{l^{\mu}}{l^2+i\epsilon}
    \end{split}
\end{align*}
to integrate out the $z$-dependence. Finally, we use the identity
 \begin{align*}
\int_a ^b dx f(x)\frac{\partial^n}{\partial x^n}\delta(x-y)=(-1)^n \frac{\partial^n}{\partial x^n}f(x)\bigg |_{x=y},
 \end{align*}
 to evaluate the integrals over the Fourier conjugates. After applying these steps, we arrive at Eqs.~\eqref{yuck} and \eqref{polyuck}.

The leading-twist contributions to the scalar amplitudes in Eq.~\eqref{expltensordecomp} are
\begin{equation}
    \begin{split}
        \mathcal{H}_1(\bomega, \vartheta, t) & = {2}\sum_{n=2,4,6}^{\infty}\bomega^{n}\int_{-1}^1dxx^{n-1}H(x,\vartheta/\bomega,t),
        \\   \mathcal{H}_2(\bomega, \vartheta, t) & = \frac{2\bar{Q}^2}{\pq}\sum_{n=2,4,6}^{\infty}\bomega^{n}\int_{-1}^1dxx^{n-1}\bigg[H(x,\vartheta/\bomega,t) -\frac{2}{n}\big[H(x,\vartheta/\bomega,t)+ E(x,\vartheta/\bomega,t)  \big]\bigg],
        \\ \mathcal{H}_{3}(\bomega, \vartheta, t) & = \frac{2\bar{Q}^2}{\pq}\sum_{n=2,4,6}^{\infty}\bomega^{n}\frac{2}{n}\int_{-1}^1dxx^{n-1}\big[H(x,\vartheta/\bomega,t)+ E(x,\vartheta/\bomega,t)  \big],
       \\  \mathcal{E}_1(\bomega, \vartheta, t) & = 2\sum_{n=2,4,6}^{\infty}\bomega^{n}\int_{-1}^1dxx^{n-1}E(x,\vartheta/\bomega,t),
       \\    \mathcal{E}_2(\bomega, \vartheta, t) & = \frac{2\bar{Q}^2}{\pq}\sum_{n=2,4,6}^{\infty}\bomega^{n}\int_{-1}^1dxx^{n-1}E(x,\vartheta/\bomega,t) ,
       \\ \tilde{\mathcal{H}}_1(\bomega, \vartheta, t) & = 2\sum_{n=2,4,6}^{\infty}\bomega^{n-1}\int_{-1}^1dxx^{n-2}\tilde{H}(x,\vartheta/\bomega,t),
       \\  \tilde{\mathcal{E}}_1(\bomega, \vartheta, t) & = 2\sum_{n=2,4,6}^{\infty}\bomega^{n-1}\int_{-1}^1dxx^{n-2}\tilde{E}(x,\vartheta/\bomega,t),
        \\  \tilde{\mathcal{H}}_2(\bomega, \vartheta, t) & = -2\sum_{n=2,4,6}^{\infty}\frac{n}{n+1}\bomega^{n-1}\int_{-1}^1dxx^{n-2}\tilde{H}(x,\vartheta/\bomega,t),
        \\  \tilde{\mathcal{E}}_2(\bomega, \vartheta, t) & =-2\sum_{n=2,4,6}^{\infty}\frac{n}{n+1}\bomega^{n-1}\int_{-1}^1dxx^{n-2}\tilde{E}(x,\vartheta/\bomega,t),
       \\\mathcal{K}_i(\bomega,\vartheta,t) & = 0, \quad \text{for all }i.
       \label{loamplitudes}
    \end{split}
\end{equation}

\onecolumngrid

\section{Feynman-Hellmann} \label{sec:appendixFH}

Starting with the $\lambda_1\lambda_2$ terms of Eq.~\eqref{pertpropexpansion}, we have
\begin{equation}
    \begin{split}
         & \int _{0} ^{\tau}d\tau _1 \int _{0}^{\tau_1} d\tau_2 \<N(\mathbf{p}')|V_1(\tau_1)V_2(\tau_2)|Y(\mathbf{p})\>+ \big (V_1\leftrightarrow V_2 \big)
        = \int _{0} ^{\tau}d\tau_1 \int _{0}^{\tau_1} d\tau_2\<N(\mathbf{p}')| e^{H_{\text{QCD}}\tau_1}
        \\ & \times \int d^3 x_1 (e^{i\mathbf{q}_1\cdot\mathbf{x}_1}+e^{-i\mathbf{q}_1\cdot\mathbf{x}_1})j_3(\mathbf{x}_1) e^{H_{\text{QCD}}(\tau_2-\tau_1)}
          \int d^3 x_2 (e^{i\mathbf{q}_2\cdot\mathbf{x}_2}+e^{-i\mathbf{q}_2\cdot\mathbf{x}_2})j_3(\mathbf{x}_2) e^{-H_{\text{QCD}}\tau_2}|Y(\mathbf{p})\> + \big (\mathbf{q}_1\leftrightarrow \mathbf{q}_2 \big).
          \label{matelemexp}
    \end{split}
\end{equation}
(Note that we use the \emph{unperturbed} time-evolution operator here, since, as in all perturbation theory, the matrix element at each order is calculated for zero-coupling.)

Next, after inserting a complete set of states, Eq.~\ref{matelemexp} becomes
\begin{equation}
    \begin{split}
        & \sum_{X}  \int\frac{d^3p_X}{(2\pi)^3}\frac{1}{2E_X(\mathbf{p}_X)}\int _{0} ^{\tau}d\tau_1 \int _{0}^{\tau_1} d\tau_2\<N(\mathbf{p}')| e^{H_{\text{QCD}}\tau_1}\int d^3 x_1 (e^{i\mathbf{q}_1\cdot\mathbf{x}_1}+e^{-i\mathbf{q}_1\cdot\mathbf{x}_1})j_3(\mathbf{x}_1) e^{H_{\text{QCD}}(\tau_2-\tau_1)}
        |X(\mathbf{p}_X)\>  
        \\ & \times \int d^3 x_2 (e^{i\mathbf{q}_2\cdot\mathbf{x}_2}+e^{-i\mathbf{q}_2\cdot\mathbf{x}_2})\<X(\mathbf{p}_X)|j_3(\mathbf{x}_2) e^{-H_{\text{QCD}}\tau_2}|Y(\mathbf{p})\> + \big (\mathbf{q}_1\leftrightarrow \mathbf{q}_2 \big)
        \\ & =  \sum_{X}  \int\frac{d^3p_X}{(2\pi)^3}\frac{1}{2E_X(\mathbf{p}_X)}\int _{0} ^{\tau}d\tau_1 \int _{0}^{\tau_1}d\tau_2e^{-(E_X(\mathbf{p}_X)-E_N(\mathbf{p}'))\tau_1}e^{-(E_X(\mathbf{p}_X)-E_Y(\mathbf{p}))\tau_2}
        \\ & \times\int d^3 x_1 (e^{i\mathbf{q}_1\cdot\mathbf{x}_1}+e^{-i\mathbf{q}_1\cdot\mathbf{x}_1})\<N(\mathbf{p}')| j_3(\mathbf{x}_1) 
        |X(\mathbf{p}_X)\>   \int d^3 x_2 (e^{i\mathbf{q}_2\cdot\mathbf{x}_2}+e^{-i\mathbf{q}_2\cdot\mathbf{x}_2})\<X(\mathbf{p}_X)|j_3(\mathbf{x}_2) |Y(\mathbf{p})\> 
        \\ & + \big (\mathbf{q}_1\leftrightarrow \mathbf{q}_2 \big)
          \label{matelemexp2}.
    \end{split}
\end{equation}
Focusing solely on the Euclidean time-dependence for a moment, we see that, if $E_Y(\mathbf{p})=E_N(\mathbf{p}')$, then
\begin{equation}
     \int _{0} ^{\tau}d\tau_1 \int _{0}^{\tau_1} d\tau_2 e^{-(E_X(\mathbf{p}_X)-E_N(\mathbf{p}'))\tau_1}e^{(E_X(\mathbf{p}_X)-E_Y(\mathbf{p}))\tau_2} =\frac{1}{E_X(\mathbf{p}_X)-E_N(\mathbf{p}')}\bigg ( \tau + \frac{e^{-(E_X(\mathbf{p}_X)-E_N(\mathbf{p}'))\tau}}{E_X(\mathbf{p}_X)-E_N(\mathbf{p}')} \bigg).
     \label{tauenhanced}
\end{equation}
And if $E_Y(\mathbf{p})\neq E_N(\mathbf{p}')$,
\begin{equation}
   \begin{split}
        & \int _{0} ^{\tau}d\tau_1 \int _{0}^{\tau_1} d\tau_2 e^{-(E_X(\mathbf{p}_X)-E_N(\mathbf{p}'))\tau_1}e^{(E_X(\mathbf{p}_X)-E_Y(\mathbf{p}))\tau_2}
    \\ & = \frac{1}{E_X(\mathbf{p}_X)-E_Y(\mathbf{p})}\bigg (\frac{e^{-(E_X(\mathbf{p}_X)-E_N(\mathbf{p}'))\tau} -1}{E_X(\mathbf{p}_X)-E_N(\mathbf{p}')}-\frac{e^{-(E_Y(\mathbf{p})-E_N(\mathbf{p}'))\tau} -1}{E_Y(\mathbf{p})-E_N(\mathbf{p}')}  \bigg ).
    \label{intercept}
   \end{split}
\end{equation}

Because of our choice of perturbing potential, the only values the source momentum can take are $\mathbf{p}=\mathbf{p}+n\mathbf{q}_1+m\mathbf{q}_2$ for $m,n\in \mathbb{Z}$ at order $\mathcal{O}(\lambda^{m+n})$. As we stated before, we choose our kinematics so that $|\mathbf{p}|\leq|\mathbf{p}+n\mathbf{q}_1+m\mathbf{q}_2|$. Therefore, for any state in the nucleon spectrum $X$ and any momentum $\mathbf{q}=\mathbf{p}'+n\mathbf{q}_1+m\mathbf{q}_2$, we must have $E_X(\mathbf{q})\geq E_N(\mathbf{p}')$.

This ensures two things: (1) that the exponentials in Eqs.~\ref{tauenhanced} and \ref{intercept} are decaying, and (2) that if $E_Y(\mathbf{p})=E_N(\mathbf{p}')$, then $Y=N$, and hence we have \textit{ground state saturation of the source}.

Therefore, 
\begin{equation}
    \begin{split}
         & \int _{0} ^{\tau}d\tau _1 \int _{0}^{\tau_1} d\tau_2 \<N(\mathbf{p}')|V_1(\tau_1)V_2(\tau_2)|Y(\mathbf{p})\>+ \big (V_1\leftrightarrow V_2 \big)= \tau\sum_{X}  \int\frac{d^3p_X}{(2\pi)^3}\frac{1}{2E_X(\mathbf{p}_X)}\frac{1}{E_X(\mathbf{p}_X)-E_N(\mathbf{p}')}
        \\ & \times\int d^3 x_1 (e^{i\mathbf{q}_1\cdot\mathbf{x}_1}+e^{-i\mathbf{q}_1\cdot\mathbf{x}_1})\<N(\mathbf{p}')| j_3(\mathbf{x}_1) 
        |X(\mathbf{p}_X)\>   \int d^3 x_2 (e^{i\mathbf{q}_2\cdot\mathbf{x}_2}+e^{-i\mathbf{q}_2\cdot\mathbf{x}_2})\<X(\mathbf{p}_X)|j_3(\mathbf{x}_2) |N(\mathbf{p})\> 
        \\ & + \big (\mathbf{q}_1\leftrightarrow \mathbf{q}_2 \big) + \Big[\text{exponentially decaying in $\tau$} \Big] + \Big [\text{constant in $\tau$}  \Big].
          \label{matelemexp3}
    \end{split}
\end{equation}
The exponentially decaying terms will be heavily suppressed for $\tau\gg a$ compared to the purely linear in $\tau$ terms and the constant. Therefore, we will neglect these. For the moment we neglect the term that is constant in $\tau$; however, we will consider this in our fit to the lattice data.

From translational invariance of the current,
\begin{equation*}
    j_3(\mathbf{x})=e^{-i\hat{\mathbf{P}}\cdot\mathbf{x}}j_3(0)e^{i\hat{\mathbf{P}}\cdot\mathbf{x}},
\end{equation*}
and hence Eq.~\ref{matelemexp3} becomes
\begin{equation}
    \begin{split}
         &  \int _{0} ^{\tau}d\tau _1 \int _{0}^{\tau_1} d\tau_2 \<N(\mathbf{p}')|V_1(\tau_1)V_2(\tau_2)|Y(\mathbf{p})\>+ \big (V_1\leftrightarrow V_2 \big)= 
         \\ & \tau\sum_{X} \int\frac{d^3p_X}{(2\pi)^3}\frac{1}{2E_X(\mathbf{p}_X)}\frac{\<N(\mathbf{p}')| j_3(0) |X(\mathbf{p}_X)\>  \<X(\mathbf{p}_X)| j_3(0)|N(\mathbf{p})\>}{E_X(\mathbf{p}_X)-E_N(\mathbf{p}')}\Delta_{12} + \big (\mathbf{q}_1\leftrightarrow \mathbf{q}_2 \big),
          \label{matelemexp4}
    \end{split}
\end{equation}
where
\begin{equation}
\begin{split}
  \Delta_{12}\equiv(2\pi)^6\Big[\delta^{(3)}(\mathbf{p}'-\mathbf{q}_1-\mathbf{p}_X)+\delta^{(3)}(\mathbf{p}'+\mathbf{q}_1-\mathbf{p}_X)\Big]\Big[\delta^{(3)}(\mathbf{p}-\mathbf{q}_2-\mathbf{p}_X)+\delta^{(3)}(\mathbf{p}+\mathbf{q}_2-\mathbf{p}_X)\Big].
  \label{deltadef}
    \end{split}
\end{equation}
Although we have kept all the delta functions here, in our final evaluation we will only keep those that ensure $|\mathbf{p}|=|\mathbf{p}'|$, as this is the condition that allowed us to take $E_N(\mathbf{p})=E_N(\mathbf{p}')$.

It is convenient to define the operator
\begin{equation*}
    \hat{\mathcal{O}}(\mathbf{p},\mathbf{q})\equiv  \sum_{X} \frac{1}{2E_X(\mathbf{p}+\mathbf{q})}\frac{ j_3(0) |X(\mathbf{p}+\mathbf{q})\>  \<X(\mathbf{p}+\mathbf{q})| j_3(0)}{E_X(\mathbf{p}+\mathbf{q})-E_N(\mathbf{p})}.
\end{equation*}
Therefore, we evaluate
\begin{equation}
    \begin{split}
    & \sum_{X} \int\frac{d^3p_X}{(2\pi)^3}   (2\pi)^6\Big[\delta^{(3)}(\mathbf{p}-\mathbf{q}_1-\mathbf{p}_X)+\delta^{(3)}(\mathbf{p}'+\mathbf{q}_1-\mathbf{p}_X)\Big]\Big[\delta^{(3)}(\mathbf{p}'-\mathbf{q}_2-\mathbf{p}_X)+\delta^{(3)}(\mathbf{p}+\mathbf{q}_2-\mathbf{p}_X)\Big] 
    \\ & \times \frac{1}{2E_X(\mathbf{p}_X)}\frac{\<N(\mathbf{p}')| j_3(0) |X(\mathbf{p}_X)\>  \<X(\mathbf{p}_X)| j_3(0)|N(\mathbf{p})\>}{E_X(\mathbf{p}_X)-E_N(\mathbf{p})} + \big (\mathbf{q}_1\leftrightarrow \mathbf{q}_2 \big)
    \\ & =    \<N(\mathbf{p}')|(2\pi)^3\Big[\delta^{(3)}(\mathbf{p}-\mathbf{q}_2+\mathbf{q}_1 - \mathbf{p}') \hat{\mathcal{O}}(\mathbf{p}',-\mathbf{q}_1)+\delta^{(3)}(\mathbf{p}-\mathbf{q}_2-\mathbf{q}_1 - \mathbf{p}') \hat{\mathcal{O}}(\mathbf{p}',\mathbf{q}_1)\\ &+\delta^{(3)}(\mathbf{p}+\mathbf{q}_2+\mathbf{q}_1 - \mathbf{p}') \hat{\mathcal{O}}(\mathbf{p}',-\mathbf{q}_1)
    +\delta^{(3)}(\mathbf{p}+\mathbf{q}_2-\mathbf{q}_1 - \mathbf{p}') \hat{\mathcal{O}}(\mathbf{p}',\mathbf{q}_1)\Big]|N(\mathbf{p})\> + \big (\mathbf{q}_1\leftrightarrow \mathbf{q}_2 \big).
    \label{deltaeval}
    \end{split}
\end{equation}
Since $|\mathbf{p}'| = |\mathbf{p}'+\mathbf{q}_1-\mathbf{q}_2|$, the only terms to survive are $$\delta^{(3)}(\mathbf{p}+\mathbf{q}_2-\mathbf{q}_1 - \mathbf{p}') \hat{\mathcal{O}}(\mathbf{p}',\mathbf{q}_1), \quad \text{and}\quad \delta^{(3)}(\mathbf{p}-\mathbf{q}_1+\mathbf{q}_2 - \mathbf{p}') \hat{\mathcal{O}}(\mathbf{p}',-\mathbf{q}_2).$$
Inserting this into Eq.~\ref{propcomboexpansion}, we have
\begin{equation}
    \begin{split}
         R_{\lambda} (\tau, \mathbf{p}')
       & \overset{\tau\gg a}{\simeq} 4\lambda^2\tau\int \frac{d^3p'}{(2\pi)^3}\frac{A_N^{\lambda}(\mathbf{p}')}{2E_N(\mathbf{p})}\<N(\mathbf{p}')|\Big (2\pi)^3[\delta^{(3)}(\mathbf{p}+\mathbf{q}_2-\mathbf{q}_1 - \mathbf{p}') \hat{\mathcal{O}}(\mathbf{p}',\mathbf{q}_1)
        \\ &+ \delta^{(3)}(\mathbf{p}-\mathbf{q}_1+\mathbf{q}_2 - \mathbf{p}')\hat{\mathcal{O}}(\mathbf{p}',-\mathbf{q}_2) \Big]|N(\mathbf{p})\> + \lambda^2 \mathcal{C} + \mathcal{O}(\lambda^4),
        \label{propcomboexpansion2}
    \end{split}
\end{equation}
where $\mathcal{C}$ is constant in $\lambda$ and $\tau$, obtained from Eq.~\ref{intercept}.

Noting that the OFCA for a single quark flavour and unit charge can be expressed as
\begin{equation*}
    \begin{split}
      T^{33}(\mathbf{p}', \mathbf{q}, \mathbf{q}') = \<N(\mathbf{p}')|\hat{\mathcal{O}}(\mathbf{p}',\mathbf{q})|N(\mathbf{p})\>+\<N(\mathbf{p}')|\hat{\mathcal{O}}(\mathbf{p}',-\mathbf{q}')|N(\mathbf{p})\>,
    \end{split}
\end{equation*}
equation \ref{propcomboexpansion2} becomes
\begin{equation}
    \begin{split}
         R_{\lambda} (\tau, \mathbf{p}')
       & \overset{\tau\gg a}{\simeq} \frac{2\lambda^2\tau}{E_N(\mathbf{p}')} T^{33}( \mathbf{p}', \mathbf{q}_1, \mathbf{q}_2) + \lambda^2 \mathcal{C}+ \mathcal{O}(\lambda^4),
        \label{propcomboexpansion3}
    \end{split}
\end{equation}
where we have used the fact that $A_N^{\lambda}(\mathbf{p}')= 1 +\mathcal{O}(\lambda)$ at most, but once again odd powers of $\lambda$ vanish as a consequence of our combination of propagators, Eq.~\ref{combocorr}.

\twocolumngrid

\bibliography{main}

\end{document}